\documentstyle[preprint,prc,eqsecnum,aps,epsf]{revtex}

\tightenlines   %for single spacing in preprint

\begin{document}
\draft

\def\lsim{\thinspace{\hbox to 8pt{\raise -5pt\hbox{$\sim$}\hss{$<$}}}\thinspace}
\def\rsim{\thinspace{\hbox to 8pt{\raise -5pt\hbox{$\sim$}\hss{$>$}}}\thinspace}

\title{Nucleon Scattering  from Very Light Nuclei:  \\
 Intermediate Energy Expansions for Transition Potentials and 
  Breakup Processes}

\author{ Ch.~Elster}
\address{
Institute of Nuclear and Particle Physics,  and
Department of Physics, \\ Ohio University, Athens, OH 45701}
    
\author{ W. Gl\"ockle}
\address{
 Institute for Theoretical Physics II, Ruhr-University Bochum,
D-44780 Bochum, Germany.}

\vspace{10mm}

\date{\today}

\maketitle

\begin{abstract}
Optical potentials for elastic p-d scattering and the coupled processes
p+$^3$He $\rightarrow$ p+$^3$He and p+$^3$He $\rightarrow$ d+d are derived 
in the Faddeev-Yakubovsky framework with special emphasis on leading
order terms, which are
 expected to be valid at intermediate energies. In addition,
equations for the fragmentations $^3$He(p,ppp)n and $^3$He(p,pp)d
are derived within the same framework. Again leading order terms
for intermediate energies are considered.
\end{abstract}

\vspace{10mm}


\pagebreak

%****************************************************************************
 
\narrowtext
 
%******************************************************************

\section{Introduction}

\hspace*{10mm}
Optical potentials have a long tradition in describing scattering of
protons and neutrons from composite targets.
At intermediate energies their theoretical 
formulation is often based on approaches like the Watson \cite{Watson}, 
the KMT formulation \cite{KMT} or the spectator expansion 
\cite{spect,bolle,med2}.
The fundamental idea from which  those approaches  start is the
grouping of the scattering process for a nucleon hitting a nuclear
 target
into rescatterings of various orders. In lowest order the projectile
interacts with one target nucleon, which has a momentum
distribution according to a mean field generated by the residual nucleus.
In second order the struck target nucleon rescatters from a second target
nucleon, which therefore participates actively in the scattering
process  and is thus no longer `hidden'
in the mean field of the target.
This more and more complex scenario can be formulated
in a general fashion in the so called spectator expansion, 
but in practice only the lowest order 
processes have been numerically
realized. In the most elaborate case this leads to the so called
full-folding model \cite{ffrc,brieva,ffce}. Here, for constructing 
the optical potential the fully-off-shell nucleon-nucleon (NN) t-matrix
is convoluted with the single nucleon density matrix of the target. 
Antisymmetrization is kept
among the target nucleons, but the projectile nucleon is treated
as distinguishable and only the antisymmetrization between the active
nucleons is kept. This procedure of treating the antisymmetrization
is clearly approximative and there might be some flaws in the
actual realization as pointed out in the Appendix A.   However,
this picture turned out to be quite successful in describing
proton and neutron scattering from light ($^{16}$O) to heavy
($^{208}$Pb) nuclei at intermediate energies 
using a NN t-matrix derived from
a realistic NN force and single nucleon density matrices resulting
from nuclear structure calculations \cite{ffrc,brieva,ffce}.
Since the structure part faces solving the full
A-body problem, approximations have to be made. In addition,
it is not the same NN
force used in the structure part as the one acting between projectile
 and target nucleon. This can be considered as an inconsistency.
The same theoretical approach has been used to describe elastic scattering
of protons from a light nucleus like $^3$He at intermediate energies
\cite{RLan,RLc}.

\hspace*{10mm}
At present only for very light nuclei there is a chance to
eliminate this inconsistency concerning the nuclear structure 
part and treat the scattering process as well as the
structure on an equal footing, namely  based on the same 
realistic NN forces.
It is the aim of the present study to derive rigorously the optical potential
and transition potentials
for scattering of nucleons from the two lightest nuclei,  the deuteron 
and $^3$He within the framework of the Faddeev-Yakubovsky equations.
In this framework, antisymmetry between all nucleons is treated
correctly.
To exhibit differences as well as similarities to  the
spectator expansion, we give a brief sketch of the latter scheme in
Appendix~A. 

\hspace*{10mm}
 In Section II we consider the proton-deuteron (p-d) system and in 
Section III the more complicated
p-$^3$He system.  In the latter case it appears natural to generalize
the  optical potential for elastic p-$^3$He scattering to a potential
matrix corresponding to a coupled set of equations describing
 scattering in both two-body fragmentation channels, p-$^3$He and d-d.
An additional interest lies in the
study of the various breakup processes
in p-$^3$He scattering into 3 fragments, proton-proton-deuteron (ppd), 
and into 4 fragments, pppn. 
For these reactions a wealth of experimental information
is available \cite{brudat,jr}. 
However, these data are conventionally analyzed in
PWIA or DWIA only, which does not correspond to a microscopic
four particle theory based on NN forces.
Three- and four-body fragmentation processes 
 will be described in Section IV.
Throughout the whole article we only consider the strong interaction
when describing protons and neglect Coulomb forces. In a practical
calculation the Coulomb interaction between the projectile proton and
the charge distribution of the nucleus can be included in a
straightforward manner at least in the elastic channel \cite{RLc,Coul1}.

\hspace*{10mm}
  In all cases to be discussed, we emphasize
intermediate energies where
the optical potential expressions and the breakup amplitudes
can be expected to simplify. Presently, in the
four-body system these simpler expressions appear to be the only ones 
which can be realized in practical applications. 
Nevertheless, they carry important
information on the reaction process and can serve as testing ground for 
applying the free
NN force in such reactions and as a means to study properties
of realistic $^3$He wave functions. Appendix~B contains the leading
terms of the transition potentials between the p-$^3$He and the
d-d channels.

\hspace*{10mm}
 The four-body scattering problem has been  formulated in  some detail in the
context of the Alt-Grassberger-Sandhas (AGS) equations \cite{AGS,Wbook}. 
This formulation is basically
equivalent to the Faddeev-Yakubovsky one which we are using. 
In practice however, and especially in the context of deriving optical
potentials for p-$^3$He reactions, which is our aim here, it is not
necessarily convenient to start from the equations presented in
Ref.~\cite{AGS}. In addition, we want to keep our formulation totally
independent of assumptions like finite rank approximations to the
underlying force. For these reasons we find it adequate to set up the
coupled four-body equations our way by regarding the asymptotic behavior
of the Faddeev-Yakubovsky equations  in configuration space and to
introduce at the earliest stage the identity of the nucleons. 
For the best of our knowledge, this sort of derivation has not been
displayed before for scattering processes. For four-nucleon bound states
however, this kind of formulation has already been used \cite{4NK}.
We conclude in Section V.

\section{Elastic Proton-Deuteron Scattering}

\hspace*{10mm}
It is well known that the operator U for elastic p-d scattering
obeys the AGS equations \cite{AGS,Wbook}
\begin{equation}
 U = P G_0^{-1} + P t G_0 U . \label{eq:2.1}
\end{equation} 
Here $G_0$ is the free three-nucleon propagator, $t$ the off-shell
NN t-operator and $P$
a sum of a cyclic and an anticyclic permutation of 3 particles.
This operator equation has to be applied onto the initial channel state
\begin{equation}
|\Phi_{\bf q_0}> \equiv |\phi_d>|{\bf q_0}> \label{eq:2.2}
\end{equation} 
where $|\phi_d\rangle$ is the deuteron state and 
$|{\bf q_0}\rangle$ the momentum
eigenstate of relative motion of the projectile nucleon with respect
to the deuteron. Throughout this Section we use the standard Jacobi momenta
${\bf p}$ for a
two-body subsystem and ${\bf q}$ for the `spectator' particle.
In addition we work
in the isospin formalism and treat the nucleons as identical.

\hspace*{10mm}
The motion of the three nucleons in intermediate states is now separated
into two parts, namely one, where a pair forms a deuteron, and another
one where this pair is in the two-body continuum.
Denoting by $H_0$ the operator for the kinetic energy of the three
nucleons and by $V$ the potential operator for the NN force,
we obtain
\begin{equation}
 t G_0 \equiv V ( E+ i \varepsilon - H_0 - V) ^{-1} \equiv V G_b + V G_c
  \label{eq:2.3} 
\end{equation} 
with
\begin{equation}
G_b \equiv |\phi_d\rangle {1 \over{E+i\varepsilon - E_d- \frac {3}{4m} q^2 }}
< \phi_d|   \label{eq:2.4}
\end{equation} 
and
\begin{equation}
 G_c\equiv \int d^3 p | \phi_{\bf p}>^{(+)} {1 \over{
E+i\varepsilon -\frac {p^2}{m} -\frac {3}{4m} q^2 }} <\phi_{\bf p}^{(+)}|
 \: .  \label{eq:2.5} 
\end{equation} 
The deuteron binding energy is given by $E_d$.
The set of two-body eigenstates $|\phi_d>$ and $|\phi_{\bf p}>^{(+)}$
span the two-body Hilbert space
and have been used in Eq.~(\ref{eq:2.4}) and Eq.~(\ref{eq:2.5})
to decompose the resolvent operator of  Eq.~(\ref{eq:2.3}).
In Eqs.~(\ref{eq:2.4}) and (\ref{eq:2.5}) the denominator is 
still an operator acting on the
spectator motion through the kinetic energy operator $\frac{3}{4m} q^2$.

\hspace*{10mm}
The driving term in Eq.~(\ref{eq:2.1}) when applied onto 
the state $|\Phi_{\bf q_0}\rangle$ given in Eq.~(\ref{eq:2.2}) 
can also be written as
\begin{equation}
P G_0^{-1} |\Phi_{\bf q_0}> = P V |\Phi_{\bf q_0}> ,  \label{eq:2.6}
\end{equation}
where $V$ is acting only within the deuteron. 

\noindent
When inserting the decomposition given in Eq.~(\ref{eq:2.3}) into
Eq.~(\ref{eq:2.1}), we obtain
\begin{equation}
U|\Phi_{\bf q_0}> = P V |\Phi_{\bf q_0}> + P V G_b U|\Phi_{\bf q_0}>
    + P V G_c U |\Phi_{\bf q_0}> . \label{eq:2.7} 
\end{equation}
We now define 
\begin{equation}
U_c |\Phi_{\bf q_0}> = PV|\Phi_{\bf q_0}> + P V G_c U_c 
   |\Phi_{\bf q_0}> , \label{eq:2.8}
\end{equation}
and consequently can rewrite Eq.~(\ref{eq:2.7}) as
\begin{equation}
U | \Phi_{\bf q_0}> = U_c |\Phi_{\bf q_0}> + 
 \int d^3 q U_c | \Phi _{\bf q}>
g_0({\bf q}) < \Phi_{\bf q} | U | \Phi_{\bf q_0}> \: . \label{eq:2.9}
\end{equation}
Here Eq.~(\ref{eq:2.4}) has been written out explicitly as
\begin{equation}
G_b = \int d^3 q |\phi_d > |{\bf q}> {1 \over{ E+i\varepsilon
- E_d-\frac {3}{4m}q^2 }}<\phi_d|<{\bf q}|\equiv \int d^3 q|\Phi_{\bf q}>
g_0({\bf q})<\Phi_{\bf q}| \: .  \label{eq:2.10}
\end{equation}
With this preparation Eq.~(\ref{eq:2.9}) can be cast into the 
closed form of an integral
equation by projecting onto the state $\langle \Phi_{\bf q '}|$ 

\begin{equation}
<\Phi_{\bf q'}|U|\Phi_{\bf q_0}>=<\Phi_{\bf q'}|U_c|\Phi_{\bf q_0}>
+\int d^3 q < \Phi _{\bf q'} |U_c | \Phi_{\bf q}> g_0({\bf q})
<\Phi_{\bf q}|U|\Phi_{\bf q_0}>  \label{eq:2.11}
\end{equation}

\noindent
This is the desired integral equation for  the elastic p-d scattering
amplitude. After a partial wave decomposition it becomes
one-dimensional.  The driving term 
\begin{equation}
<\Phi _{\bf q' }|U_c|\Phi_{\bf q}> \equiv V_{\bf q' \bf q} \label{eq:2.12}
\end{equation}
can be interpreted as  optical potential $V_{\bf q' \bf q}$.
The question of interest is, whether at intermediate
energies (say above 100 MeV)
one can expect that the integral equation, Eq.~(\ref{eq:2.8}), 
can be solved by iteration
and moreover whether the very first few terms will be sufficient.
Assuming this  to be correct we expand
\begin{equation}
<\Phi_{\bf q'}|U_c|\Phi_{\bf q}>\approx <\Phi_{\bf q'}|P V |
\Phi_{\bf q}> + <\Phi_{\bf q'} | P V G_c P V | \Phi_{\bf q}> +\cdots 
  \label{eq:2.13} 
\end{equation}
In order to proceed further, we need to be more specific and identify a  
two-body subsystem, e.g.
we choose the pair (23) to be the deuteron.
Then the potential $V$ becomes $V\equiv V_{23}$. The permutation
operator $P$ is given by
\begin{equation}
P\equiv P_{12} P_{23} + P_{13} P_{23}, \label{eq:2.13a}
\end{equation}
and the channel state by
\begin{equation}
|\Phi_{\bf q}>\equiv | \Phi_{\bf q}>_1\equiv |\phi_d>_{23}|{\bf q>}_1.
\label{eq:2.14}
\end{equation}
The first term of Eq.~(\ref{eq:2.13}) is now explicitly given as
\begin{equation}
<\Phi_{\bf q'} | P V | \Phi_{\bf q}> =_1 <\Phi_{\bf q'} |
V_{31}|\Phi_{\bf q}>_2 + \;  _1<\Phi_{\bf q'}|V_{12}|\Phi_{\bf q}>_3 \: ,
\label{eq:2.15}
\end{equation}
where 
\begin{eqnarray}
| \Phi_{\bf q}>_2 & \equiv & P_{12} P_{23} |\Phi_{\bf q}>_1\equiv |\phi_d>_
{31}|\bf q >_2   \\ \nonumber
| \Phi_{\bf q}>_3 & \equiv & P_{13}P_{23}|\Phi_{\bf q}>_1\equiv |\phi_d>_
{12}|\bf q >_3  .  \label{eq:2.17}
\end{eqnarray}
With this, one easily finds 
\begin{equation}
<\Phi_{\bf q'}|P V | \Phi_{\bf q}>=_1<\Phi_{\bf q'}|V_{13}
(1-P_{13})|\Phi_{\bf q}>_2 .    \label{eq:2.18} 
\end{equation}

\noindent
The potential $V_{13}$, the NN force, should only contribute in two-body
states which obey the Pauli principle. Therefore,
it has to be represented by antisymmetrized states and thus
\begin{equation}
V_{13}(1-P_{13}) = 2 V_{13}  .  \label{eq:2.19}
\end{equation}
It follows  that
\begin{equation}
<\Phi_{\bf q'}|P V | \Phi_{\bf q}> = 2 \;\; _1<\Phi_{\bf q'}|
V_{13}|\Phi_{\bf q}>_2 \: .   \label{eq:2.20}
\end{equation}

\noindent
This is the well known single particle exchange process where the
deuterons in the bracket states are formed out of nucleons
(23) and (13), respectively.  If the spectator momenta $ \bf q$ and
$\bf q'$ are larger than the typical momenta inside the deuteron
this matrix element is small and will not be the leading contribution
to the optical potential.

\hspace*{10mm}
Let us now consider the second term in Eq.~(\ref{eq:2.13}).
It is convenient to introduce a t-operator $t_c$ defined as
\begin{equation}
V G_c\equiv t_c G_0 .   \label{eq:2.21}
\end{equation}
From its very definition it is related to the full $t$-operator
by
\begin{equation}
t_c=t-V|\phi_d>g_0<\phi_d|G_0^{-1}     \label{eq:2.22}
\end{equation}
and obeys the Lippmann Schwinger equation
\begin{equation}
t_c = V-V|\phi_d><\phi_d|+t_c G_0 V .  \label{eq:2.23}
\end{equation}
 With the help of Eq.~(\ref{eq:2.22}) and Eq.~(\ref{eq:2.13a}) 
the second term in Eq.~(\ref{eq:2.13}) can be expressed as 
\begin{eqnarray}
< \Phi_{\bf q'}|PVG_cPV|\Phi_{\bf q}>&=&<\Phi_{\bf q'}|
Pt_cG_0PV|\Phi_{\bf q}> \\ \nonumber
&=&<\Phi_{\bf q'}| t_c(31)(1-P_{13}) G_0 V_{23}|\Phi_{\bf q}> \\
\nonumber
&+ &<\Phi_{\bf q'}|t_c(21)(1-P_{12})G_0 V_{23}|\Phi_{\bf q}> . \label{eq:2.25}
\end{eqnarray}
\noindent
If the momentum ${\bf q}$ would be on shell, i.e.
$|{\bf q}|=|{\bf q_0}|$, 
then we would have $G_0V_{23}|\Phi_{\bf q}\rangle=|\Phi_{\bf q}\rangle$
and we would end up with a simple
form, namely that nucleon $1$ interacts with the constituents 
of the deuteron (nucleons $2$ and $3$) through $t_c$. 
However, the expression is more complicated.
The free propagator is given as
\begin{equation}
G_0 = {1 \over{ E_d- p^2/m}}-G_0(E-E_d- \frac {3}{4m} q^2) {1 \over{E_d
-p^2/m}} . \label{eq:2.26}
\end{equation}
Therefore, 
\begin{equation}
G_0 V_{23}|\Phi_{\bf q}> = | \Phi_{\bf q}> - G_0 g_0^{-1}|
\Phi_{\bf q }> , \label{eq:2.27}
\end{equation}
so that the second term in Eq.~(\ref{eq:2.13}) is given by
\begin{eqnarray}
<\Phi_{\bf q'}|PVG_cPV|\Phi_{\bf q}> &=&  \nonumber \\
 <\Phi_{\bf q'}|t_c(13)(1-P_{13})(1-G_0g_0^{-1})|\Phi_{\bf q}>
&+&<\Phi_{\bf q'}|t_c(12)(1-P_{12})(1-G_0g_0^{-1})|\Phi_{\bf q}> \: .
\label{eq:2.28}
\end{eqnarray}

\noindent
This expression can be reduced by replacing $(1-P_{ij})$ by $2$
and noticing that because of the antisymmetry of the deuteron states
the two matrix elements are the same.
\noindent
This resulting form for the optical potential has some similarity
to the often used `$t\rho$' form as discussed in  Appendix~A.
It will be interesting to perform numerical studies in order to see
quantitatively the validity of the truncation in Eq.~(\ref{eq:2.13})
 for the optical
potential and to compare this expression with the more 
standard `$t\rho$' form.
Such investigations are planned.

\section{Two-Body Fragmentations in Proton-$^3$He Scattering}

\hspace*{10mm}
Four-nucleon scattering has not yet been numerically
mastered in a rigorous and
general manner similar to the scattering of three nucleons upon each 
other \cite{WPRep}. 
However, at intermediate energies there is a chance, that specific 
approximations can be theoretically 
justified and systematically controlled in a numerical realization.  
This can open interesting
insight into the reaction itself, the use of the free (not modified
through the presence of the nuclear medium)  NN interaction in such
processes
 as well as into the
properties of the three-nucleon bound state. Our approach
is a rigorous framework and has been successfully used for the
$\alpha$-particle bound state \cite{Kam1,Kam2}.

\hspace*{10mm}
As is well known the fully antisymmetric scattering state for
p-$^3$He scattering, $\Psi^{(+)}$, obeys the homogeneous integral
equation 
\begin{equation}
\Psi^{(+)} = G_0 \sum_{ij} V_{ij} \Psi^{(+)},    \label{eq:3.1}
\end{equation}
where $G_0$ is the free four-nucleon (4N) propagator 
and $V_{ij}$ represents the NN force between particles $i$ and $j$. 
As a first
step one usually decomposes $\Psi^{(+)}$ into 6 Faddeev components
\begin{equation}
\Psi^{(+)} =\sum_{ij} \psi_{ij}   \label{eq:3.2}
\end{equation}
with 
\begin{equation}
\psi_{ij}\equiv G_0 V_{ij} \Psi^{(+)}= G_0 V_{ij} \sum_{k l} \psi_{kl}.
\label{eq:3.3}
\end{equation}
When introducing the off-shell NN t-matrix $t_{ij}$, as already
defined in Section  II,
one arrives at 6 coupled Faddeev equations
\begin{equation}
\psi_{ij} = G_0 t_{ij}\sum_{kl\ne ij}\psi_{kl} .  \label{eq:3.4}
\end{equation}

\hspace*{10mm}
In the following we immediately make use of the identity of the 
four particles.
This simplifies the notation significantly and it appears not to have been
outlaid before in the literature in this particular form. 
It also serves to clearly define our notation. Without loss of
generality we consider the scattering of particle 4 from a target
consisting of the subcluster built from particles 1, 2 and 3.

\noindent
Let us consider the Faddeev component
\begin{equation}
\psi_{12} = G_0 t_{12} ( \psi_{23} +\psi_{31} +
\psi_{24} + \psi_{14} + \psi_{34}),  \label{eq:3.5}
\end{equation}
which is then split into 3 Yakubovsky components, namely
\begin{equation}
\psi_1\equiv G_0 t_{12}(\psi_{23}+\psi_{31}) ,  \label{eq:3.6}
\end{equation}
\begin{equation}
P_{34}\psi_1 = - G_0 t_{12} ( \psi_{24} + \psi_{14}) ,  \label{eq:3.7}
\end{equation}
and
\begin{equation}
\psi_2\equiv G_0 t_{12} \psi_{34} .  \label{eq:3.8}
\end{equation}
To arrive at Eq.~(\ref{eq:3.7}) we used the antisymmetry of 
$\Psi^{(+)}$ and
 the definition of the Faddeev components as given in 
Eq.~(\ref{eq:3.3}). This gives
\begin{equation}
\psi_{12}= ( 1-P_{34}) \psi_1 + \psi_2 .  \label{eq:3.9}
\end{equation}
With the same reasoning one has
\begin{equation}
\psi_{23}+\psi_{31} = P\psi_{12},   \label{eq:3.10}
\end{equation}
where the permutation operator $P$ is given in Eq.~(\ref{eq:2.13a}).
Thus Eq.~(\ref{eq:3.6}) can be written as 
\begin{equation}
\psi_1 = G_0 t_{12} P \psi_{12} = G_0 t_{12} P ( (1-P_{34})\psi_1 + \psi_2).
\label{eq:3.11}
\end{equation}

\noindent
The decisive step to describe p-$^3$He scattering is to sum up all 
pair forces within the 3-body subcluster
of particles 1,2 and 3. In order to achieve this, we rewrite 
Eq.~(\ref{eq:3.11}) as
\begin{equation}
(1-G_0 t_{12}P) \psi_1=G_0 t_{12} P (-P_{34}\psi_1 + \psi_2). \label{eq:3.12}
\end{equation}

\noindent
Now the left hand side alone has a nontrivial solution, which is related
to the incoming channel state via
\begin{equation}
(1-G_0 t_{12}P)\Phi^F = 0  \label{eq:3.13}
\end{equation}
with  
\begin{equation}
\Phi^F = |\phi^F(123)\rangle|{\bf u}\rangle_4 \: . \label{eq:3.14}
\end{equation}

\noindent
Here $|\phi^F\rangle$ denotes the Faddeev component to the target state,
namely the 3N bound state $|\phi\rangle$,
\begin{equation}
|\phi>\equiv ( 1+P) |\phi^F>.   \label{eq:3.15}
\end{equation}
The momentum eigenstate of the projectile, $|{\bf u}>_4$, is
described by an
appropriate Jacobi momentum $\bf u$. 
With these definitions we can write Eq.~(\ref{eq:3.12})
as
\begin{equation}
\psi_1 =\Phi^F +(1-G_0 t_{12}P)^{-1} G_0 t_{12} P (-P_{34}\psi_1+\psi_2).
\label{eq:3.16}
\end{equation}
Defining now
\begin{equation}
(1-G_0 t_{12} P)^{-1} G_0 t_{12} P\equiv G_0 T P , \label{eq:3.17}
\end{equation}
 which is  equivalent to introducing the 3-body operator
\begin{equation}
T=t_{12} + t_{12}PG_0 T ,  \label{eq:3.18}
\end{equation}
one obtains
\begin{equation}
\psi_1 = \Phi^F + G_0 T P ( -P_{34} \psi_1 + \psi_2) . \label{eq:3.19}
\end{equation}

\noindent
The three-body operator $T$ given in 
Eq.~(\ref{eq:3.18}) is defined by an off-shell Faddeev equation 
for the three-body subsystem composed
of particles 1,2, and 3, and depends parametrically
on the kinetic energy of the fourth particle.  
The above Eq.~(\ref{eq:3.19}) is the first one of the 
Yakubovsky-equations.

\hspace*{10mm}
  Let us now consider the second Yakubovsky component from
Eq.~(\ref{eq:3.8}). With
 the help of the permutation operator
\begin{equation}
\tilde P\equiv P_{13}P_{24}  \label{eq:3.20}
\end{equation}
we can write
\begin{equation}
\psi_2 = G_0 t_{12} \tilde P\psi_{12} = G_0 t_{12} \tilde P ((1-P_{34})
 \psi_1 + \psi_2) . \label{eq:3.21}
\end{equation}

\noindent
Next, we sum up the pair forces in the two noninteracting two-body
subsystems (12) and (34), representing two deuterons, 
to infinite order. To do this, we rewrite
Eq.~(\ref{eq:3.21}) as
\begin{equation}
(1-G_0 t \tilde P) \psi_2=G_0 t \tilde P(1-P_{34}) \psi_1   \label{eq:3:22}
\end{equation}
and solve for $\psi_2$.
In the two-deuteron channels there are no ingoing waves, therefore
the nontrivial solution to the left hand side alone should not be
added and we obtain
\begin{equation}
\psi_2 = (1-G_0 t \tilde P)^{-1} G_0 t \tilde P(1-P_{34}) \psi_1
= G_0 \tilde T\tilde P ( 1-P_{34}) \psi_1  \label{eq:3.23}
\end{equation}
with 
\begin{equation}
\tilde T=t_{12}+ t_{12} \tilde P G_0  \tilde T.   \label{eq:3.24}
\end{equation}
The above equation, Eq.~(\ref{eq:3.24}), defines the T-operator for the
two two-particle subclusters, which
interact only internally but not with each other.
The desired second Yakubovsky equation is given by Eq.~(\ref{eq:3.23}).
For the sake of
completeness we give also the final expression for the total
scattering state
\begin{equation}
\Psi^{(+)} = ( 1+P-P_{34}P + \tilde P) ((1-P_{34})\psi_1+\psi_2)
\label{eq:3.25}
\end{equation}

\hspace*{10mm}
In order to define an optical potential for the scattering of
 a nucleon from $^3$He, we have to separate 
in Eq.~(\ref{eq:3.19}) the propagation
in the $^3$He subcluster channel from the propagation, where the three 
nucleons 1,2, and 3 are in intermediate scattering states.  In
order to achieve this, 
we have to reconsider the definition given in Eq.~(\ref{eq:3.17})
in the following way:
\begin{eqnarray}
( 1-G_0 t_{12} P)^{-1} G_0 t_{12} P  &= & 
( 1-G V_{12} P)^{-1} G V_{12} P   \nonumber \\
 &= &( E-H_0 -V_{12}(1+P))^{-1} V_{12} P \nonumber \\
 &\equiv & G_0 T P . \label{eq:3.26}
\end{eqnarray}
Here we introduced the resolvent operator 
\begin{equation}
G\equiv ( E+i\varepsilon - H_0 -V_{12}). \label{eq:3.27}
\end{equation}
In the four-body system the convenient `odd man out'
notation conventionally  used in the three-body system is not applicable 
and what there appears 
`natural' as choice of the arbitrarily singled out pair, namely
$1\equiv(23)$, is not being applied to the four-body system. Instead,
we have the pair (12) as the `first' pair.
Again, the specific choice of the `first' pair is irrelevant and just
 a matter of convenience.

\hspace*{10mm}
By the very definition of the target state $|\phi \rangle$ and its 
Faddeev component $|\phi^F \rangle$
given in Eq.~(\ref{eq:3.15}) one has
\begin{equation}
(E_{He} - h_0(123) - V_{12}(1+P))|\phi^F>=0   \label{eq:3.28}
\end{equation}
and
\begin{equation}
<\phi |(E_{He} -h_0(123)-V_{12}(1+P))=0. \label{eq:3.29}
\end{equation}
Here the 3N binding energy is given by $E_{He}$,
 and $h_0(123)$ stands for
the internal kinetic energy of the nucleons 1, 2 and 3.  The above two 
equations indicate that there are different left and right eigenvectors, 
and thus we obtain
\begin{equation}
(E+i\epsilon - H_0-V_{12}(1+P))^{-1} = |\phi^F > g_0 
{1 \over{<\phi|\phi^F>}} <\phi| + continuum \:\: , \label{eq:3.30}
\end{equation}
where 
\begin{equation}
 g_0 = {1 \over{E+i\epsilon-E_{He} - \frac {2}{3m}{\bf u}^2}}
  \label{eq:3.31}
\end{equation}
is now the single particle propagator in the $^3$He subcluster channel,
while in Section II it was used for describing 
the propagation in the deuteron channel.
Using Eq.~(\ref{eq:3.30}) we now can cleanly separate this propagation 
from the rest
\begin{equation}
G_0 T P \equiv |\phi^F> g_0{1 \over{<\phi|\phi^F>}} <\phi|
  V_{12}P + G_0 T^c P . \label{eq:3.32}
\end{equation}

\noindent
Inserted into Eq.~(\ref{eq:3.19}) we arrive at the final expression for the
Yakubovsky component $\psi_1$: 
\begin{equation}
\psi_1 = \Phi^F + |\phi^F > g_0 {1 \over{<\phi|\phi^F>}}<\phi| V_{12}P| 
( -P_{34}\psi_1 +\psi_2)>+G_0 T^c P ( -P_{34}\psi_1 +\psi_2). \label{eq:3.33}
\end{equation}

\noindent
Obviously, the second term on the right hand side provides an
 asymptotically purely outgoing wave carrying the elastic amplitude
 for p-$^3$He scattering
\begin{equation}
M\equiv {1 \over{<\phi|\phi^F>}}<\Phi|V_{12}P(-P_{34}\psi_1 +\psi_2)>.
\label{eq:3.34}
\end{equation}
Here \mbox{$\langle \Phi|=\langle \phi|\langle \bf u|$} is 
the on-shell channel state. 
This form is clearly not the standard one and we sketch briefly the
link to the more familiar expression. To achieve this, we first rewrite
Eq.~(\ref{eq:3.34}) as
\begin{eqnarray}
<\phi|\phi^F>M &=&< \Phi | V_{12}P ( -P_{34}\psi_1 + \psi_2)> \nonumber \\
&=&<\Phi| ( E-H_0 -V_{12})|( -P_{34}\psi_1 + \psi_2)>  \nonumber \\
&=& <\Phi| ( E-H_0 -V_{12})|(\psi_{12}-\psi_1)> . \label{eq:3.35}
\end{eqnarray}
The second equality corresponds to Eq.~(\ref{eq:3.29}) 
for $\langle \Phi|$ as introduced above and the 4N kinetic energy $H_0$.
The third equality is due to Eq.~(\ref{eq:3.9}). 
Inserting now the definitions of
 $\psi_{12}$ and $\psi_1$, Eqs.~(\ref{eq:3.3}) and (\ref{eq:3.11}), 
we obtain 
\begin{eqnarray}
<\phi|\phi^F> M & = &  < \Phi |(E-H_0 -V_{12}) G_0 V_{12} \Psi^{(+)}>
 -< \Phi |(E-H_0 -V_{12}) G_0 t P G_0 V_{12} | \Psi^{(+)}> \nonumber \\
& = & <\Phi | V_{12} | \Psi^{(+)} > -<\Phi| V_{12}G_0 V_{12} |\Psi^{(+)}>-
  <\Phi| V_{12}PG_0 V_{12} | \Psi^{(+)}> .  \label{eq:3.36}
\end{eqnarray}
\noindent
Now we use Eq.~(\ref{eq:3.1}) and get together with the
explicit definition of the permutation operator as given in
Eq.~(\ref{eq:2.13a})
\begin{equation}
<\phi|\phi^F> M  = <\Phi | V_{12} G_0 ( V_{14}+V_{24}+V_{34})|\Psi^{(+)}>.
 \label{eq:3.37}
\end{equation}

\noindent
Due to the antisymmetry of $\Phi$ with respect to particles 1, 2 and 3 
and of \mbox{$(V_{14}+V_{24}+V_{34})|\Psi^{(+)} \rangle$} this 
can be rewritten as
\begin{eqnarray}
<\phi|\phi^F> M &=&1/3 <\Phi| ( V_{12}+V_{23}+V_{31})G_0 (V_{14}+V_{24}+
V_{34})|\Psi^{(+)}>  \nonumber \\
&=&1/3<\Phi|V_{14}+V_{24}+V_{34}|\Psi^{(+)}> . \label{eq:3.38}
\end{eqnarray}
In the last equality we took advantage of 
$\langle \Phi| ( V_{12}+V_{23}+V_{31})G_0 = \langle \Phi|$. 
Finally we note that $\langle \phi|\phi^F \rangle =1/3$ 
for a normalized state $|\phi \rangle$.
This concludes our verification that the amplitude $M$ 
given in Eq.~(\ref{eq:3.34}) is indeed the desired
elastic scattering amplitude for p-$^3$He scattering. 

\hspace*{10mm}
Starting from Eq.~(\ref{eq:3.33}) and Eq.~(\ref{eq:3.23}) one can already 
work out the optical potential formalism for p-$^3$He scattering. 
We do not want to do this here but rather in addition 
separate off the propagation of two deuterons in 
Eq.~(\ref{eq:3.23}). This allows to derive 
coupled equations for p-$^3$He and 
d-d scattering.  Very likely this will accelerate the convergence of
the expansions of the resulting `optical' transition potentials.  In
order to achieve the separation of the d-d channel we proceed analogously
to the separation of the p-$^3$He channel.
First we note that in analogy to Eq.~(3.26)
\begin{equation}
G_0 \tilde T\tilde P= ( E-H_0 -V_{12}(1+\tilde P))^{-1} V_{12} \tilde P. 
\label{eq:3.39}
\end{equation}
Now we need to introduce the analog to `Faddeev' components
for the d-d channel, which are somewhat unfamiliar. The two
uncorrelated deuterons $\phi_d ( 12)$ and $\phi_d(34)$ obey the
Schr\"odinger equation 
\begin{equation}
( h_0(12)+V_{12}-E_d +h_0(34)+V_{34}-
E_d)\phi_d(12)\phi_d(34)=0 , \label{eq:3.40}
\end{equation}
where $h_0(ij)$ are internal kinetic energies only.
Let us call
\begin{equation}
  \phi_{dd}\equiv |\phi_d>_{12}|\phi_d>_{34}. \label{eq:3.41}
\end{equation}
In view of Eq.~(\ref{eq:3.39}) we can rewrite Eq.~(\ref{eq:3.40}) as
\begin{equation}
<\phi_{dd}| ( 2 E_d -h_0(12)-h_0(34)-V_{12}(1+\tilde P))=0 .
\label{eq:3.42}
\end{equation}
The integral form of Eqs.~(\ref{eq:3.42}) or (\ref{eq:3.40}) then reads 
\begin{equation}
\phi_{dd} = {1 \over{2 E_d-h_0(12)-h_0(34)}}
(V_{12}+V_{34})\phi_{dd} \: .  \label{eq:3.43}
\end{equation}

\noindent
The state $\phi_{dd}$ can now be decomposed in a Faddeev-like fashion as
\begin{equation}
\phi_{dd}=\phi_{dd}^{F,12} + \phi_{dd}^{F,34}. \label{eq:3.44}
\end{equation}
 with
\begin{equation}
\phi_{dd}^F\equiv \phi_{dd}^{F,12} \equiv 
{1 \over{2 E_d -h_0(12)-h_0(34)}} V_{12} \phi_{dd} \label{eq:3.45}
\end{equation}
and 
\begin{equation}
\phi_{dd}^{F,34} = \tilde P \phi_{dd}^F. \label{eq:3.46}
\end{equation}
It follows that 
\begin{equation}
(2 E_d-h_0(12)-h_0(34)- V_{12}(1+\tilde P))\phi_{dd}^F=0 .
 \label{eq:3.47}
\end{equation}

\noindent
Using Eqs.~(\ref{eq:3.42}) and (\ref{eq:3.47}) we can separate the
d-d channel from the rest
\begin{equation}
G_0 \tilde T\tilde P \equiv |\phi_{dd}^F> g^{dd}_0 {1 \over{ <\phi_{dd}
|\phi_{dd}^F>}}<\phi_{dd}|V_{12}\tilde P + G_0 \tilde T^c \tilde P.
 \label{eq:3.48}
\end{equation}
Here the propagator in the d-d channel is explicitly given by 
\begin{equation}
g^{dd}_0\equiv{1 \over {E-2 E_d +i\epsilon -\frac {1}{2m}{\bf v}^2}},
\label{eq:3.49}
\end{equation}
where $\bf v$ is a suitable Jacobi momentum for the relative
motion of the two deuterons.

\noindent
Insertion of Eq.~(\ref{eq:3.48}) into Eq.~(\ref{eq:3.23}) 
gives for  the second Yakubovsky equation
\begin{equation}
\psi_2 = |\phi_{dd}^F> g^{dd}_0 {1 \over{<\phi_{dd}|\phi_{dd}^F>}}
<\phi_{dd}|V_{12}\tilde P(1-P_{34})\psi_1 + G_0 \tilde T^c \tilde P
 ( 1-P_{34}) \psi_1. \label{eq:3.50}
\end{equation}

\noindent
The asymptotic form for the d-d channel follows in a similar fashion  
as for Eq.~(\ref{eq:3.33}) and the
resulting transition amplitude into the d-d channel is given by
\begin{equation}
M_{dd} \equiv {1 \over{ < \phi_{dd}|\phi_{dd}^F>}}
 <\Phi_{dd}|V_{12}\tilde P(1-P_{34}|\psi_1>,  \label{eq:3.51}
\end{equation}
where $\Phi_{dd}$ denotes the  d-d channel state
\begin{equation}
\Phi_{dd} = | \phi_{dd}>|\bf v> \label{eq:3.52}
\end{equation}

\noindent
Again, this is not the standard form for the transition amplitude.
We want to briefly sketch the relation to the standard form. The basics
steps are very similar to the ones leading to the elastic amplitude
given in Eq.~(\ref{eq:3.38}). We start from
\begin{eqnarray}
<\phi_{dd}|\phi_{dd}^F> M_{dd} & =& 
<\Phi_{dd}|V_{12}\tilde P(1-P_{34}|\psi_1> \nonumber \\
& = & <\Phi_{dd}|(E-H_0 -V_{12}) ( 1-P_{34})|\psi_1> \nonumber \\
& = & <\Phi_{dd}|(E-H_0 -V_{12})|\psi_{12}-\psi_2> \nonumber \\
& = & <\Phi_{dd}|(E-H_0 -V_{12})G_0 V_{12}|\Psi^{(+)}>-
<\Phi_{dd}|(E-H_0 -V_{12})G_0
t\tilde P G_0 V_{12}|\Psi^{(+)}> \nonumber \\
& = & <\Phi_{dd}|V_{12}|\Psi^{(+)}> -
<\Phi_{dd}|V_{12}G_0 V_{12} |\Psi^{(+)}>-<\Phi_{dd}|V_{12}
 \tilde PG_0 V_{12}|\Psi^{(+)}> \nonumber \\
& = & <\Phi_{dd}|V_{12}G_0(V_{13}+V_{14}+V_{23}+V_{24})|\Psi^{(+)}>
      \nonumber \\
& = & 1/2 <\Phi_{dd}|(V_{12}+V_{34})G_0(V_{13}+V_{14}+V_{23}+V_{24})|
\Psi^{(+)}> \nonumber \\
& = & 1/2<\Phi_{dd}|(V_{13}+V_{14}+V_{23}+V_{24})|\Psi^{(+)}>.\label{eq:3.53}
\end{eqnarray}
Since $\langle \phi_{dd}|\phi_{dd}^F \rangle =1/2$ 
we confirm the standard form.

\vspace{4mm}
\hspace*{10mm}
 Now we are ready to derive coupled equations for the 
p-$^3$He and d-d channels starting from
Eqs.~(\ref{eq:3.33}) and (\ref{eq:3.50}). 
As a first step we introduce coupled amplitudes for general
Jacobi momenta ${\bf u}$ and ${\bf v}$ corresponding to 
the driving terms either in the p-$^3$He or the d-d channel,
\begin{eqnarray}
\psi_{1,\bf u}^c &=& \Phi_{\bf u}^F + G_0 T^c P ( -P_{34}\psi_{1,\bf u}^c +
 \psi_{2,\bf u}^c)  \nonumber \\
\psi_{2,\bf u}^c &=& G_0 \tilde T^c \tilde P(1-P_{34}) \psi_{1,\bf u}^c
\label{eq:3.55}
\end{eqnarray}
and
\begin{eqnarray}
\psi_{1,\bf v}^{d}&=&G_0 T^c P ( -P_{34}\psi_{1,\bf v}^{d} +
\psi_{2,\bf v}^{d})  \nonumber \\
\psi_{2,\bf v}^{d}&=& \Phi_{dd,\bf v}^{F}+G_0 \tilde T^c \tilde P(1-P_{34}) 
\psi_{1,\bf v}^{d} \label{eq:3.57}
\end{eqnarray}
Here $\Phi_{dd,{\bf v}}^F=\phi_{dd}^F |{\bf v} \rangle$ is 
the `Faddeev' component in the d-d channel together with the
relative momentum eigenstate $|{\bf v}\rangle$ of the two deuterons.
The second set of equations, Eqs.~(\ref{eq:3.57}),
is necessary to derive coupled channel equations.
With the above definitions it follows from Eqs.~(\ref{eq:3.33}) and 
(\ref{eq:3.50})
\begin{eqnarray}
\psi_1 & = & \psi_{1,\bf u_0}^c + \int d^3 u \psi_{1,{\bf u}}^c g_0(u)
{1 \over <\phi|\phi^F>}<\Phi_{\bf u}|V_{12} P ( -P_{34}\psi_1 +\psi_2)>
\nonumber \\
& +& \int d^3 v \psi_{1,\bf v}^{d} g^{dd}_0 ({\bf v})
 {1 \over <\phi_{dd}| \phi_{dd}^F>}<\Phi_{\bf v}^{dd}|V_{12} 
 \tilde P(1-P_{34})|\psi_1> \label{eq:3.58}
\end{eqnarray}
and
\begin{eqnarray}
\psi_2 & = & \psi_{2,\bf u_0}^c +\int d^3 u \psi_{2,\bf u}^c g_0({\bf u})
{1 \over <\phi|\phi^F>}<\Phi_{\bf u}|V_{12} P ( -P_{34}\psi_1 +\psi_2)>
\nonumber \\
&+&\int d^3 v \psi_{2,\bf v}^{d} g_0^{dd} ({\bf v}) {1 \over{ <\phi_{dd}|
\phi_{dd}^F>}}<\Phi_{\bf v}^{dd}|V_{12}\tilde P (1-P_{34})|\psi_1>.
 \label{eq:3.59}
\end{eqnarray}

\noindent
Here we denote the initial on-shell relative momentum by ${\bf u_0}$
in contrast to the previous use.
The right hand sides contain the transition amplitudes from the
initial channel p-$^3$He to the channels p-$^3$He and d-d. 
Starting from these equations we can easily obtain the following
set of coupled equations for half-shell transition amplitudes:
\begin{eqnarray}
 & & <\Phi_{\bf u'}|V_{12}P|(-P_{34}\psi_1 +\psi_2)>  \:\: = \:\: 
  <\Phi_{\bf u'}|V_{12}P|(-P_{34} \psi_{1,\bf u_0}^c +\psi_{2,\bf u_0}^c)>
   \nonumber \\
&+&  \int d^3 u  <\Phi_{\bf u'}|V_{12} P|(-P_{34}
\psi_{1,\bf u}^c +\psi_{2,\bf u}^c)>g_0({\bf u}){1 \over{ <\phi|\phi^F>}}
<\Phi_{\bf u}|V_{12}P|(-P_{34}\psi_1 +\psi_2)> \nonumber \\
&+&  \int d^3 v  <\Phi_{\bf u'}|V_{12}P|(-P_{34}\psi_{1,\bf v}^{d}
+\psi_{2,\bf v}^{d})>g_0^{dd}({\bf v}){1 \over{ <\phi_{dd}| 
\phi_{dd}^F>}}<\Phi_{\bf v}^{dd}|V_{12}\tilde P(1-P_{34})|\psi_1>
   \label{eq:3.60}
\end{eqnarray}
and
\begin{eqnarray}
 & &<\Phi_{\bf v'}^{dd}|V_{12}\tilde P(1-P_{34})|\psi_1> \:\: =\:\:  
<\Phi_{\bf v'}^{dd}|V_{12}\tilde P(1-P_{34})|\psi_{1,\bf u_0}^c> \nonumber \\
&+& \int d^3 u <\Phi_{\bf v'}^{dd}|V_{12}\tilde P 
 (1-P_{34})|\psi_{1,\bf u}^c> g_0({\bf u}) {1 \over <\phi|\phi^F>}
 <\Phi_{\bf u}|V_{12} P ( -P_{34}\psi_1 +\psi_2)> \nonumber \\
&+&\int d^3 v<\Phi_{\bf v'}^{dd}|V_{12}\tilde P(1-P_{34})|
\psi_{1,\bf v}^{d}> g_0^{dd}({\bf v}){1 \over{ <\phi_{dd}|
\phi_{dd}^F>}}<\Phi_{\bf v}^{dd}|V_{12}\tilde P(1-P_{34})|\psi_1> 
 \label{eq:3.61}
\end{eqnarray}

\noindent
Defining transition potentials as

\begin{eqnarray}
V_{\bf u' \bf u} &= &<\Phi_{\bf u'}|V_{12} P|(-P_{34}\psi_{1,\bf u}^c +
\psi_{2,\bf u}^c)>  \nonumber \\
V_{\bf v' \bf v}&=&<\Phi_{\bf v'}^{dd}|V_{12} \tilde P(1-P_{34})|
\psi_{1,\bf v}^{d}>  \nonumber \\
V_{\bf u' \bf v}&=&<\Phi_{\bf u'}|V_{12} P|(-P_{34}\psi_{1,\bf v}^{d}
 +\psi_{2,\bf v}^{d})>  \nonumber \\
V_{\bf v' \bf u}&=&<\Phi_{\bf v'}^{dd}|V_{12} \tilde P(1-P_{34})
 |\psi_{1,\bf u}^c)> , \label{eq:3.62}
\end{eqnarray}
we can rewrite Eqs.(\ref{eq:3.60}) and (\ref{eq:3.61}) in a more
transparent form:
\begin{eqnarray}
 <\Phi_{\bf u'}|V_{12}P|(-P_{34}\psi_1 +\psi_2)> & =& \: V_{\bf u'u_0}
 \nonumber \\ 
&+&  \int d^3 u V_{\bf u'u} g_0({\bf u}){1 \over{ <\phi|\phi^F>}}
<\Phi_{\bf u}|V_{12}P|(-P_{34}\psi_1 +\psi_2)> \nonumber \\
&+&  \int d^3 v V_{\bf u'v} g_0^{dd}({\bf v}){1 \over{ <\phi_{dd}| 
\phi_{dd}^F>}}<\Phi_{\bf v}^{dd}|V_{12}\tilde P(1-P_{34})|\psi_1>
   \label{eq:3.63}
\end{eqnarray}
\begin{eqnarray}
<\Phi_{\bf v'}^{dd}|V_{12}\tilde P(1-P_{34})|\psi_1>& =&  \: 
  V_{\bf v',u_0} \nonumber \\
&+& \int d^3 u V_{\bf v'u} g_0({\bf u}) {1 \over {<\phi|\phi^F>}}
 <\Phi_{\bf u}|V_{12} P| ( -P_{34}\psi_1 +\psi_2)> \nonumber \\
&+& \int d^3 v V_{\bf v'v} g_0^{dd}({\bf v}){1 \over{ <\phi_{dd}|
\phi_{dd}^F>}}<\Phi_{\bf v}^{dd}|V_{12}\tilde P(1-P_{34})|\psi_1> 
 \label{eq:3.64}
\end{eqnarray}
These are the coupled integral equations for the 
two-body fragmentation channels
p-$^3$He and d-d and constitute the main result in this study.
The solutions lead directly to the elastic amplitude $M$ and the transition
amplitude into the d-d channel $M_{dd}$ as given in Eqs.~(\ref{eq:3.34}) and
(\ref{eq:3.51}).

\vspace{2mm}
\hspace*{10mm}
The remaining task is to determine the amplitudes $\psi_i^c$ and
$\psi_i^d$ as given 
in Eqs.~(\ref{eq:3.55}) and (\ref{eq:3.57}).
The idea and hope is that at high enough energies this set of amplitudes
can be obtained by iteration and that only the first few terms will contribute.
To be explicit let us consider the second order iteration of those
equations, which is given by

\begin{eqnarray}
\left(\matrix{\psi_{1,\bf u}^c \cr
    \psi_{2,\bf u}^c \cr}\right)&=&\left(\matrix{\Phi_{\bf u}^F\cr
                                                      0\cr}\right)
+\left(\matrix{G_0T^cP(-P_{34})\Phi_{\bf u}^F \cr
                  G_0\tilde T^c \tilde P(1-P_{34})\Phi_{\bf u}^F\cr}\right)
\nonumber \\
&+&\left(\matrix{G_0T^cP(-P_{34})G_0T^cP(-P_{34})\Phi_{\bf u}^F
+G_0T^cPG_0\tilde T^c\tilde P(1-P_{34})\Phi_{\bf u}^F\cr
 G_0\tilde T^c\tilde P(1-P_{34})G_0T^cP(-P_{34})\Phi_{\bf u}^F\cr}\right)
\label{eq:3.66}
\end{eqnarray}
and
\begin{eqnarray}
\left(\matrix{\psi_{1,\bf v}^{d} \cr
      \psi_{2,\bf v}^{d} \cr}\right)&=&\left(\matrix{0\cr
                                     \Phi_{dd,\bf v}^F\cr}\right)
+ \left(\matrix{ G_0T^cP \Phi_{dd,\bf v}^F\cr
                    0\cr}\right) \nonumber \\
&+&\left(\matrix{G_0T^cP(-P_{34})G_0T^cP \Phi_{dd,\bf v}^F\cr
  G_0\tilde T^c\tilde P(1-P_{34})G_0T^cP \Phi_{dd,\bf v}^F\cr}\right)
\label{eq:3.67}
\end{eqnarray}

\noindent
Inserting this result into Eq.~(\ref{eq:3.62}) we find for  example 
the transition potential
for p-$^3$He to p-$^3$He scattering in this order of approximation
\begin{eqnarray}
V_{\bf u',\bf u}&=&<\Phi_{\bf u'}|V_{12}P|(-P_{34})|\Phi_{\bf u}^F>
\nonumber \\
&+&<\Phi_{\bf u'}|V_{12}P(-P_{34})G_0T^cP(-P_{34}) +
  V_{12}P G_0\tilde T^c \tilde P(1-P_{34})| \Phi_{\bf u}^F> \nonumber \\
&+&<\Phi_{\bf u'}|V_{12}P(-P_{34})G_0T^cP(-P_{34})G_0T^c P(-P_{34})
  + V_{12}P (-P_{34})G_0T^cPG_0\tilde T^c\tilde P(1-P_{34})
  \nonumber \\
&+&V_{12}P G_0\tilde T^c\tilde P(1-P_{34})G_0T^cP(-P_{34})|\Phi_{\bf u}^F>.
\label{eq:3.68}
\end{eqnarray}
The other transition potentials can be obtained in a similar fashion
and are given in Appendix~B.

\hspace*{10mm}
In order to evaluate the transition potentials we have to 
determine  $T^c$ and $\tilde T^c$.
According to the definitions given in Eqs.~(\ref{eq:3.32}) and 
(\ref{eq:3.48}) as well as the defining equations (\ref{eq:3.18}) 
and (\ref{eq:3.24}) for $T$ and $\tilde T$, we obtain after some algebra
\begin{equation}
G_0T^c = \left[1- |\phi^F>{1 \over{<\phi|\phi^F>}}<\phi| \right] G_0t_{12}
+ G_0T^c PG_0t_{12} \label{eq:3.69}
\end{equation}
and
\begin{equation}
G_0 \tilde T^c = \left[ 1- |\phi_{dd}^F>{1 \over{<\phi_{dd}|\phi_{dd}^F>}}
<\phi_{dd}| \right] G_0t_{12}
+ G_0 \tilde T^c\tilde P G_0t_{12} . \label{eq:3.70}
\end{equation}
As a consequence of the uniqueness of these equations and also 
from their very definition 
(the latter one  requiring some algebraic manipulations) 
one can conclude that
\begin{equation}
<\phi |G_0T^c =0 \label{eq:3.71}
\end{equation}
and
\begin{equation}
<\phi_{dd}|G_0\tilde T^c =0 \label{eq:3.72}
\end{equation}
This excludes intermediate propagation of $^3$He or d-d states
in the amplitudes given in Eqs.~(\ref{eq:3.66}) and ({\ref{eq:3.67}).
Of course, this should be the case.
From solving 3N Faddeev equations at intermediate energies \cite{WPRep},
we know that the multiple scattering series converges, except for the
$^3$He quantum numbers $J^\pi = 1/2^+$. This divergence, however,
is due to the very existence
of $^3$He. Due to the special driving terms in Eqs.~(\ref{eq:3.69})
and (\ref{eq:3.70}) this divergence is removed. 
Therefore, there is a good reason to assume that
Eqs.~(\ref{eq:3.69}) and (\ref{eq:3.70}) can be successfully iterated and 
that very low orders are sufficient.
With the abbreviations                  
\begin{equation}
\Lambda \equiv 1-|\phi^F>{1 \over{<\phi|\phi^F>}}<\phi | \label{eq:3.73}
\end{equation}
and
\begin{equation}
\Lambda_{dd} \equiv 1-|\phi_{dd}^F>{1 \over{<\phi_{dd}|\phi_{dd}^F>}}
<\phi_{dd} | \label{eq:3.75}
\end{equation}
we consequently approximate $G_0 T^c$ and $G_0 \tilde T^c$ to second
order in $t_{12}$ as
\begin{equation}
G_0 T^c\rightarrow\Lambda G_0 t_{12} + \Lambda G_0 t_{12} P G_0 t_{12} 
\label{eq:3.74}
\end{equation}
and
\begin{equation}
G_0 \tilde T^c\rightarrow\Lambda_{dd} G_0 t_{12} + \Lambda_{dd} G_0 t_{12}
\tilde  P G_0 t_{12}.
\label{eq:3.76}
\end{equation}

\noindent
When inserting these expressions into  Eq.~(\ref{eq:3.68}) we obtain
for the transition potential $V_{\bf u'u}$

\begin{eqnarray}
V_{\bf u'\bf u}& =& <\Phi_{\bf u'}|V_{12}P(-P_{34})|\Phi_{\bf u}^F> 
   \nonumber \\
&+& <\Phi_{\bf u'}|V_{12}PP_{34} \Lambda G_0 t_{12}  P P_{34}
 + V_{12}P \Lambda_{dd} G_0 t_{12}\tilde P ( 1- P_{34})|\Phi_{\bf u}^F>
\nonumber \\
&+& <\Phi_{\bf u'}|V_{12}PP_{34} \Lambda G_0 t_{12}PG_0t_{12}PP_{34}
 + V_{12}P \Lambda_{dd} G_0 t_{12} \tilde PG_0 t_{12}\tilde P( 1- P_{34})
 | \Phi_{\bf u}^F> \nonumber \\
& - & <\Phi_{\bf u'} | V_{12}PP_{34}\Lambda G_0 t_{12}PP_{34}\Lambda 
 G_0t_{12}PP_{34} + V_{12}PP_{34} \Lambda G_0
t_{12}P\Lambda_{dd} G_0 t_{12}\tilde P(1-P_{34}) \nonumber \\
& + & V_{12}P\Lambda_{dd} G_0t_{12}\tilde P(1-P_{34})\Lambda G_0
t_{12}PP_{34}|\Phi_{\bf u}^F>.
\label{eq:3.77}
\end{eqnarray}

\noindent
In the parts of $V_{\bf u' u}$ which are second order in $T^c$  and
$\tilde T^c$ we kept only the first order parts of Eqs.~(\ref{eq:3.74})
and (\ref{eq:3.76}) which are linear in $t_{12}$.

\noindent
As one of several possible simplifications in the above expression,
we note that
 $\tilde P(1-P_{34})=
(1-P_{12})\tilde P$ and $(1-P_{12})$ yields a factor of 2
upon its application to $t_{12}$.  Thus we obtain for the
transition potential
\begin{eqnarray}
V_{\bf u'\bf u} & = & <\Phi_{\bf u'}|V_{12}P(-P_{34})|\Phi_{\bf u}^F> 
  \nonumber \\
&+&<\Phi_{\bf u'}|V_{12}PP_{34} \Lambda G_0 t_{12}PP_{34}
 + 2 V_{12}P \Lambda_{dd} G_0 t_{12}\tilde P |\Phi_{\bf u}^F> \nonumber \\
&+& <\Phi_{\bf u'}|V_{12}PP_{34} \Lambda G_0 t_{12} PG_0t_{12}PP_{34}
   +2 V_{12}P \Lambda_{dd} G_0 t_{12}\tilde P G_0t_{12}\tilde P|
   \Phi_{\bf u}^F>   \nonumber \\
&-& <\Phi_{\bf u'}| V_{12}PP_{34}\Lambda G_0 t_{12}PP_{34}\Lambda G_0
 t_{12}PP_{34} 
 + 2 V_{12}PP_{34}\Lambda G_0t_{12}P\Lambda_{dd} G_0t_{12}\tilde P
  \nonumber \\
& & +2 V_{12}P\Lambda_{dd} G_0t_{12}\tilde P \Lambda G_0t_{12}PP_{34}
  |\Phi_{\bf u}^F>.
\label{eq:3.78}
\end{eqnarray}

\noindent
The evaluation of this expression requires the know-how of handling
all the separate subcluster problems: 
The $^3$He bound state and its Faddeev
 components, the deuteron and the Faddeev component to the two
 deuteron subcluster problem,
the NN t-matrices and above all the (3+1) and (2+2) subcluster kernels 
$G_0t_{12}P$ and $G_0t_{12}\tilde P$. 
In addition there is the $P_{34}$ permutation
operator to be considered.  
All those mathematical structures have been successfully handled
in previous work \cite{4NK,WPRep}, and thus the evaluation of this transition
potential should pose no additional difficulty, although it is certainly 
a computational challenge.
Similar expressions can be derived for the remaining transition potentials. 
They are given in Appendix~B.

\hspace*{10mm}
A close inspection of Eq.~(\ref{eq:3.78}) shows that most of the 
terms will likely not contribute at higher nucleon projectile energies, 
since they are
exchange terms, where the projectile momentum probes bound state wave
functions in a region where they are already very small.  
But there are of course
also `$t\rho$'- type structures, like part of the fourth term in
Eq.~(\ref{eq:3.78}), which will dominate.  We leave a more detailed
study of Eq.~(\ref{eq:3.78}) to future work. The importance 
of different terms should become apparent when being supported by numerical
realization.
Nevertheless we would like to emphasize that the presented approach
is systematic. It relies on a mathematically and physically well
founded basis.  Scattering and bound state structures
are treated on the same footing, and antisymmetrization 
is included fully. The internal validity of our approach can then be checked
numerically by adding further terms of the expansions. 
There will be no free parameters in the calculation, 
 once a certain NN force has been selected.

\hspace*{10mm}
At a later stage it might be also of interest to study the properties
of those transitions potentials with respect to their spin-dependencies
and locality versus nonlocality. It would be a surprise if they would
have much in common with the often used phenomenological
Wood-Saxon type expressions.

\section{Three- and Four-Body Fragmentations in Proton-$^3$He Scattering}

\hspace*{10mm}
As alternative to an algebraic derivation as pursued in the
previous section, we would like to arrive at 
the set of coupled integral equations for the breakup process
as given in the Yakubovsky scheme starting  from
a graphical approach.
The resulting equations are exact and can be derived rigorously
in an algebraic manner.

\hspace*{10mm}
The complete breakup process initiated for example
by nucleon number 4 striking a $^3$He target composed
of nucleons 1,2, and 3 is given by the infinite sequence of processes
depicted in Fig.~1.
Here  $\phi$ to the very right of each diagram
represents the $^3$He target ground state and the dashed lines stand
for NN interactions.
Each diagram has to be read from right to left and each has 
to start with an interaction
between the projectile and one of the constituents of the target.
After this initial interaction of the projectile, arbitrary interactions
between all four particles have to occur. Clearly, this comprises
all possible interactions and intermediate free propagations of the
four nucleons.
The superscript on the breakup operator $U^{(4)}_0$ indicates
that here particle 4 has been singled out as projectile.
In order to achieve full antisymmetrization, only the interchange between
the projectile and the single target nucleons have to be considered,
since the target ground state is assumed to be already antisymmetrized. 
Thus, the fully antisymmetrized breakup operator $U_0$ is given by the
set of diagrams in Fig.~2. There, the terms representing nucleons 1, 2
or 3 as projectiles enter with a negative sign, since only one transposition
is necessary to interchange nucleon 4 with one of them.

\hspace*{10mm}
For each of the three terms $U_0^{(k)}, k=1,2,3$ in Fig.~2 
expansions corresponding to $U_0^{(4)}$ as given in Fig.~1
can be written down.
The fully antisymmetrized breakup operator $U_0$ can 
be decomposed into 6 Faddeev components according to the last pair 
interaction on the left,
\begin{equation}
U_0= \sum_{i<j} U^0_{ij}.  \label{eq:4.3}
\end{equation}
By inspection of the various Born series represented in
Fig.~2, one can read off the different Faddeev components. For example,
the component $U^0_{34}$ is given by
amplitude
\begin{equation}
U^0_{34} = V_{34}(\Phi_4 -\Phi_3) +  V_{34} G_0 \sum_{i<j} U^0_{ij}.
  \label{eq:4.4}
\end{equation}
Here $|\Phi_4 \rangle =|\phi(123) \rangle |{\bf u}\rangle_4$ and 
$|\Phi_3 \rangle  = P_{34} |\Phi_4 \rangle$.
As usual one can sum up $V_{34}$ to infinite order into $t_{34}$ :
\begin{eqnarray}
U^0_{34}& =& t_{34}(\Phi_4-\Phi_3)+t_{34} G_0 \sum_{i<j,ij\neq 34} U^0_{ij}
    \nonumber \\
 &=& t_{34}(\Phi_4 -\Phi_3) + t_{34}G_0 (U^0_{24}+U^0_{23})
    \nonumber \\
 & & + t_{34}G_0 (U^0_{14}+U^0_{13}) + t_{34}G_0 U^0_{12}.
\label{eq:4.5}
\end{eqnarray}
In the second equality we already group the terms in a suggestive 
manner, namely such that 
the pair indices in the last three terms define two different three-body 
subclusters, namely  234 and 134, and one 2+2 fragmentation 12-34. 

\hspace*{10mm}
From here on we use the identity of the particles, which 
simplifies matters considerably and allows to restrict the discussion 
always to one type of amplitude, either from the 
partition 3+1 or from the partition 2+2. All remaining amplitudes are 
obtained by suitable permutations of the particles. Like in
the previous section we define a (3+1) Yakubovsky component  
\begin{equation}
U^0_1 \equiv t_{34}G_0 (U^0_{24}+U^0_{23}) \label{eq:4.6}
\end{equation}
and a (2+2) Yakubovsky component
\begin{equation}
U^0_2 \equiv t_{34}G_0 U^0_{12} \label{eq:4.6a}
\end{equation}
as two independent amplitudes. Using the identity of the 
particles, the (3+1) component becomes 
\begin{eqnarray}
U^0_1 &=&t_{34}G_0 (-P_{23} -P_{24}) U^0_{34} \nonumber \\  
 &=& t_{34}G_0 (-P_{23} -P_{24}) \left[  t_{34}(\Phi_4 -\Phi_3) +
   U^0_1 -P_{12}U^0_1 +U^0_2 \right] \; .
\label{eq:4.7}
\end{eqnarray}
It is convenient to introduce a permutation operator for 
particles 2,3, and 4
\begin{equation}
P_{234}=P_{23}P_{34} + P_{24}P_{34} , \label{eq:4.8}
\end{equation}
since
\begin{equation}
P_{34} U^0_{34} = - U^0_{34}. \label{eq:4.9}
\end{equation}
Then
\begin{eqnarray}
U^0_1&=& t_{34}G_0 P_{234} t_{34}(1-P_{34}) \Phi_4 \nonumber \\
  & & + t_{34}G_0 P_{234} (1-P_{12}) U^0_1 + 
 t_{34}G_0 P_{234} U^0_2.  \label{eq:4.10}
\end{eqnarray}
Analogous to Eq.~(\ref{eq:3.26}) we sum up all interactions 
in the 234 subsystem to 
infinite order and obtain
\begin{equation}
U^0_1=T G_0 P_{234} t_{34}(1-P_{34}) \Phi_4 
 + T G_0 P_{234} (-P_{12}U^0_1 + U^0_2). \label{eq:4.11}
\end{equation}
The above equation defines
the (off-shell) three-particle operator $T$  as
\begin{equation}
(1-t_{34} G_0 P_{234})^{-1} t_{34} G_0 P_{234} \equiv
  T G_0 P_{234},  \label{eq:4.12}
\end{equation}
which is equivalent to the integral equation
\begin{equation}
T= t_{34} + t_{34} G_0 P_{234} T. \label{eq:4.13}
\end{equation}
For this off-shell three-body operator T the same notation is
used as in Eq.~(\ref{eq:3.18}), 
since it represents exactly the same quantity, however it is expressed in 
different particle numbers.  Similar steps as those described above
lead to
\begin{equation}
U^0_2= \tilde T G_0 \tilde P t_{34} (1-P_{34}) \Phi_4 
  + \tilde T G_0 \tilde P (1-P_{12}) U^0_1, \label{eq:4.14}
\end{equation}
where $\tilde P$ is given in Eq.~(\ref{eq:3.20})
and
\begin{equation}
\tilde T = t_{34} + t_{34} G_0 \tilde P \tilde T . \label{eq:4.16}
\end{equation}
Again the same remark as above applies  concerning $\tilde T$.
The coupled set of Eqs.~(\ref{eq:4.11}) and
~(\ref{eq:4.14}) for $U^0_1$ and $U^0_2$ are the Yakubovsky equations 
containing (3+1) and (2+2) subcluster T-operators.
Once solved, the Yakubovsky components $U_1^0$ and $U_2^0$ are 
sufficient to generate the full breakup operator $U_0$.

\hspace*{10mm}
Summing all Faddeev components as given in Eq.~(\ref{eq:4.3}) and
taking advantage of the identity of the particles yields for 
the breakup operator
\begin{equation}
U_0 = (1+P_{234} + \tilde P - P_{12}P_{234}) U^0_{34} \label{eq:4.17}
\end{equation}
with $U^0_{34}$ given in Eq.~(\ref{eq:4.7})
\begin{equation}
U^0_{34}= t_{34}(1-P_{34}) \Phi_4 +(1-P_{12}) U^0_1 +U^0_2
\label{eq:4.18}.
\end{equation}

\noindent
At intermediate energies one can expect that the lowest order terms in 
the NN t-matrix should be sufficient.
When inserting the driving terms of Eqs.~(\ref{eq:4.13}) and (\ref{eq:4.16})
into the Yakubovsky equations
(\ref{eq:4.11}) and (\ref{eq:4.14}), we obtain in lowest order
\begin{eqnarray}
U^0_1 &\approx & t_{34} G_0 P_{234} t_{34} (1-P_{34}) \Phi_4 \nonumber \\
U^0_2 &\approx & t_{34} G_0 \tilde P t_{34} (1-P_{34}) \Phi_4.
       \label{eq:4.19}
\end{eqnarray}
Thus, in lowest order the breakup operator $U_0$ is given as
\begin{eqnarray}
U_0&=& (1+ P_{234} +\tilde P -P_{12}P_{234}) \nonumber \\
& & \left[ t_{34}(1-P_{34}) \Phi_4 + (1-P_{12}) t_{34} G_0 P_{234} t_{34}
 (1-P_{34}) \Phi_4  +t_{34}G_0 \tilde P t_{34} (1-P_{34}) \Phi_4 \right].
   \label{eq:4.20}
\end{eqnarray} 
This expression is manifestly antisymmetric in all four particles, 
since the square bracket
is separately antisymmetric in the pairs 12 and 34, as is obvious for 
the first and second term. Similarly, this can be seen for the third 
term by noting that $\tilde P t_{34}(1-P_{34}) = t_{12}(1-P_{12}) 
\tilde P$.
Now  $\tilde P$ applied onto $\Phi_4$ yields  $-\Phi_2$, 
which is antisymmetric in 34.  
The total antisymmetry  can then  be checked 
using the permutation operators in front of the square bracket.

\hspace*{10mm}
As an example for the actual evaluation of the terms in Eq.~(\ref{eq:4.20}) 
let us consider
the term of first order in the NN t-matrix. It is a simple exercise 
to show that
\begin{eqnarray}
(1+P_{234} &+& \tilde P -P_{12} P_{234})t_{34}(1-P_{34})\Phi_4 \nonumber \\
&=& t_{14}(1-P_{14})\Phi_4 +t_{24}(1-P_{24})\Phi_4 + t_{34}(1-P_{34})\Phi_4
    \nonumber \\
& & -t_{12}((1-P_{12})\Phi_2 -t_{13}(1-P_{13})\Phi_1 
    -t_{23}((1-P_{23})\Phi_2 .  \label{eq:4.20b}
\end{eqnarray}
The breakup amplitude results by acting from the left with momentum 
eigenstates for four free particles.
They can be represented by 3 Jacobi momenta suitably chosen according to 
the pair interacting in the various NN t-matrices. In addition, the various 
channel states $\Phi_i, \: i=1,4$ require adequate choices of  
Jacobi momenta.  For instance $\Phi_4$ is best
described by choosing
\begin{eqnarray}
{\bf p} & = & \frac {1}{2} ({\bf k}_1 - {\bf k}_2)\nonumber \\
{\bf q} & = & \frac {2}{3} ({\bf k}_3 - \frac {1}{2} ({\bf k}_1 + {\bf
k}_2)  \nonumber \\
{\bf r} & = & \frac {3}{4} \left( {\bf k}_4 - \frac {1}{3} ({\bf k}_1 +
 {\bf k}_2  + {\bf k}_3) \right) . \label{eq:4.21} 
\end{eqnarray}
Then
\begin{equation}
<{\bf p q r} | \Phi_4> = \delta ({\bf r} - {\bf r}_0) \phi ({\bf p q}),
\label{eq:4.23}
\end{equation}
where ${\bf r}_0$ is the initial projectile momentum (in the zero total
momentum frame) and $\phi (\bf{p q})$ stands for the 3N bound state.
For the sake of simplicity we drop spin and isospin quantum numbers.
In order to describe for instance $t_{34}$ it is convenient to introduce a 
second set of Jacobi momenta, 
which singles out the relative momentum among particles 3 and 4:
\begin{eqnarray}
{\bf p}_1 & = & \frac {1}{2} ({\bf k}_3 - {\bf k}_4) \nonumber \\
{\bf q}_1 & = & \frac {2}{3} ({\bf k}_2 - \frac {1}{2}
({\bf k}_3 + {\bf k}_4)  \nonumber \\ 
{\bf r}_1 & = & \frac {3}{4} ({\bf k}_1 - \frac {1}{3} 
 ({\bf k}_2 + {\bf k}_3 + {\bf k}_4)   .  \label{eq:4.24}
\end{eqnarray}
Then one easily derives
\begin{eqnarray}
<{\bf p}_1 {\bf q}_1 {\bf r}_1 | {\bf p q r}>  & = &  \delta ({\bf p} -
\frac {2}{3} {\bf r}_1 + \frac {1}{2} {\bf q}_1) \times \nonumber \\  
& &\delta  ({\bf q} - \frac {2}{3} {\bf p}_1 + \frac {2}{3} {\bf q}_1 + 
 \frac {4}{9}
{\bf r}_1) \delta ({\bf r} + {\bf p}_1 + \frac {1}{2} {\bf q}_1 + 
\frac {1}{3} {\bf r}_1)  \nonumber \\
& = & \delta ({\bf p}_1 + \frac {2}{3} {\bf r} - \frac{1}{2} {\bf q}) 
\delta ({\bf q}_1 + \frac {2}{3} {\bf p} + \frac {2}{3}  {\bf q} +
\frac {4}{9} {\bf r}) \times \nonumber \\
& & \delta  ({\bf r}_1 -{\bf p} + \frac {1}{2} {\bf q}+\frac {1}{3}{\bf r}). 
 \label{eq:4.25}
\end{eqnarray}
Furthermore, one has
\begin{equation}
<{\bf p}_1 {\bf q}_1 {\bf r}_1 | t_{34}|{\bf p}_1' {\bf q}_1' {\bf r}_1'>
 = t_{34} ({\bf p}_1,{\bf p}_1';E - \frac {3}{4m} q^2_1 - \frac {2}{3m}
 r^2_1) \delta ({\bf q}_1 - {\bf q}_1' ) \delta ({\bf r}_1 -
{\bf r}_1'). \label{eq:4.26}
\end{equation}
Therefore, using Eqs.~(\ref{eq:4.23}) and (\ref{eq:4.26}) one easily 
finds
\begin{eqnarray}
<{\bf p}_1 {\bf q}_1 {\bf r}_1|t_{34}(1 - P_{34}) \Phi_4 >& =& \nonumber \\
2 \: t_{34} ({\bf p}_1,-{\bf r}_0 -\frac {1}{2} {\bf q}_1 -\frac {1}{3} 
{\bf r}_1 &; & E - \frac {3}{4m} q^2_1 - \frac {2}{3m} r^2_1) 
\phi \left( \frac {2}{3} {\bf r}_1 - \frac {1}{2}  {\bf q}_1, -\frac {2}{3}
 {\bf r}_0- {\bf q}_1 - \frac {2}{3} {\bf r}_1 \right) . \label{eq:4.27}
\end{eqnarray} 
This term is maximal under the condition, that both arguments in 
$\phi$ are zero. This leads to
\begin{eqnarray}
{\bf q}_1 & = & - \frac {4}{9} {\bf r}_0 \nonumber \\ 
{\bf r}_1 &=& - \frac {1}{3}  {\bf r}_0   \label{eq:4.28}
\end{eqnarray}
Noting that ${\bf r}_0 = \frac {3}{4} {\bf k}_{lab}$, where ${\bf k}_{lab}$ is 
the projectile momentum in the laboratory system,
one easily deduces by using  Eq.~(\ref{eq:4.24}),
that the conditions  Eq.~(\ref{eq:4.28}) are equivalent to 
${\bf k}_1 = {\bf k}_2 = 0.$ 
The condition of zero momenta for the spectator nucleons
is usually called quasifree scattering.

\hspace*{10mm}
Next we consider the arguments of the NN t-matrix $t_{34}$ in 
Eq.~(\ref{eq:4.27}). Using
\begin{equation}
E = E_{He} + \frac {2}{3m}  r^2_0 = \frac {p^2_1}{m} +
\frac {3}{4m} q^2_1 + \frac {2}{3m} r^2_1 ,  \label{eq:4.29}
\end{equation}
where $E_{He}$ is the binding energy of $^3$He, 
one finds under the condition given in Eq.~(\ref{eq:4.28})
\begin{equation}
t_{34}({\bf p}_1,-\frac {2}{3} {\bf r}_0; E_{He} + 
\frac {1}{m} (\frac {2}{3} {\bf r}_0)^2) .  \label{eq:4.30}
\end{equation}
Therefore $t_{34}$ is on the energy shell except for the negative 
binding correction $E_{He}$ of the $^3$He target.

\hspace*{10mm}
Going away from the quasielastic peak one can probe 
the off-shell NN t-matrix and the $^3$He 
target wave function. At the same time however, 
the other terms in Eq.~(\ref{eq:4.20b}) will also contribute. 
Each term in Eq.~(\ref{eq:4.20b}) 
expressed in the adequate Jacobi momenta will have  a form
similar to Eq.~(\ref{eq:4.27}). Corresponding 
spectator momenta set to zero have the effect, that one term
will dominate. The other ones are suppressed, since the arguments of   
$\phi$  then differ
from zero and the t-matrices are off-shell.  Numerical studies
are required
to learn about interferences among the six terms  
if one is away from the peak conditions.
One can expect to extract interesting information on the ground state
wave function $\phi$ of $^3$He
 and the off-shell NN t-matrix. Both are nowadays readily 
accessible for realistic NN forces and 
can therefore be tested.

\hspace*{10mm}
The next step is to evaluate the terms second order in $t_{34}$ as given 
in Eq.~(\ref{eq:4.20}). This is more complicated, since it
involves the free propagator, which together with the permutation 
operators leads to logarithmic singularities. However, their treatment
is known from the three-body system \cite{WPRep}, and thus these 
terms can also be numerically determined.
We expect that with increasing energy the multiple scattering series 
should terminate quickly and therefore the reaction
as well as properties of $^3$He can be tested systematically and in a
rigorous way. 
This will be left to  future work.

\hspace*{10mm}
A final step is  the formulation of the transition to the 
three-body fragmentation channels d+p+p. Again we start by 
choosing particle 4 as projectile and display the first terms of 
the infinite series of processes in Fig.~3.
Here one has to note that the last interaction to the left cannot 
take place between particles 1 
and 2, since $V_{12}$ is already taken into account in the deuteron 
ground state wave function $\phi_d(12)$.
We recognize the same series of processes as in Fig.~1, 
except that the 
last interaction $V_{12}$ is excluded.
The antisymmetrization in the initial state leads to 
the symmetrized transition operator
\begin{equation}
U_d = U_d^{(4)} - \sum_{i=1}^{3}  U^{(i)}_d , \label{eq:4.31}
\end{equation} 
where $U^{(i)}_d$  refers to the projectile $i$. For $i=1,2$ 
there occurs just 
one term with $V_{12}$, namely $V_{12}\phi_i, \: i=1,2$. Thus we have
\begin{equation}
U_{d} = - V_{12} (\phi_1 + \phi_2) +\sum_{i<j,ij \neq 12}
U^0_{ij} \: .   \label{eq:4.32}
\end{equation}
For the on-shell process the potential $V_{12}$
can be replaced by $G^{-1}_0$. 
Finally, permutation operators can be introduced leading to
\begin{equation}
U_d = - G^{-1}_0 (P_{14} + P_{24}) \Phi_4 + (1 + P - P_{12} P)
U^0_{34}.   \label{eq:4.33}
\end{equation}
Here $U^0_{34}$ is given in terms of the two Yakubovsky amplitudes $U^0_1$
and $U^0_2$ of Eq.~(\ref{eq:4.18}).
Again the lowest order terms
in Eq.~(\ref{eq:4.19}) are expected to be sufficient at
intermediate energies.
This remains to be verified in a numerical study.

\section{Summary and Outlook}
\hspace*{10mm}
Elastic proton-deuteron scattering is formulated within the
framework of Faddeev equations to 
achieve the form of a one-body equation with an optical potential
as driving term.
The exact equation defining this
potential is assumed to be solvable at intermediate energies 
by a low order expansion in the NN t-matrix. This remains to be 
verified numerically.
The presented formulation is a systematic expansion in 
two-nucleon t-matrices and treats the antisymmetry
among all three particles exactly.
  A standard approach to derive microscopic optical potentials
at intermediate energies is the `spectator expansion' of multiple
scattering theory.  We apply this
method in the case of p-deuteron scattering in order to illustrate
similarities and differences to the treatment within the Faddeev scheme.

\hspace*{10mm}
 In the four-nucleon system we formulate
an exact set of coupled equations  for p-$^3$He and d-d scattering within
the framework of the Yakubovsky equations.
The transition potentials between those two channels are again approximated
at intermediate energies in a low order expansion in the NN t-matrix.
This expansion is systematic and antisymmetrization among all
four particles is treated exactly. The numerical realization as well as the
internal check of convergence with respect to higher order terms
should provide interesting insight into the reaction
mechanism and should lay a firm ground to test various physical assumptions,
like the application of free NN forces in such a reaction 
as well as the properties of the
$^3$He wave function as resulting from solving the Faddeev equations
based again on NN forces.

\hspace*{10mm}
   We also derived exact equations for the p-$^3$He induced three-
and four-body fragmentation processes. Here we again concentrate 
on the lowest
order terms in a NN t-matrix expansion, which we expect to be valid 
at intermediate energies.
  The present work lies a formal ground for numerical investigations,
which are planned. The expectation is that at least at intermediate energies
the four-nucleon scattering problem can be numerically controlled 
in a reliable manner, using the outlaid systematic t-matrix expansion.
At lower energies such an expansion fails, as is already
known from the 3N system \cite{WPRep}. 
In the low energy regime an exact solution of the four-nucleon
Yakubovsky equations appears not to be feasible at the present time.
Since the proposed scheme does not contain any adjustable 
parameters once a realistic
NN force has been chosen, its numerical realization and comparison
to experiment should provide interesting insight into our understanding
of such reaction processes and the properties of the
$^3$He wave function.

\vfill
\acknowledgements
This work was performed in part under the auspices of the U.S.
Department of Energy under contract No. DE-FG02-93ER40756 with
Ohio University.
 One of the authors (W.G) would like to thank
the Department of Physics and Astronomy at the Ohio University for 
offering him a Putnam Professorship. 
%=====================================================

\appendix
\section{The spectator expansion for p-d scattering}

\hspace*{10mm}
In this Appendix, the most simple system for nucleon-nucleus
scattering, the p-d system, is used to explain the steps involved in the
spectator expansion of multiple scattering theory, on which one 
model for the nucleon-nucleus optical potential is based \cite{med2}.
Since the p-d system is the simplest system for nucleon-nucleus scattering
all steps can be clearly carried out and the result can be compared to
the exact Faddeev framework of Section II. We especially want to emphasize
the treatment of the antisymmetrization and show the inherent limitation
in spectator expansion. For distinguishable particles the spectator
expansion is carried out for three particles in Ref.~\cite{monster}.

\hspace*{10mm}
In the general derivation of the optical potential for scattering of
a nucleon from a composite nucleus, the protons and neutrons are treated
as distinguishable particles. In this Appendix we will follow this
general practice. 
Numbering the two protons as particles 1 and 2 
and the neutron as particle 3, the scattering state initiated by 
proton 1 scattering from  a deuteron composed of nucleons 2 and 3 is
given by
\begin{equation}
\Psi_1^{(+)} = i \epsilon G \Phi_1 , \label{eq:A.1}
\end{equation}
where  $G$ the full resolvent operator and
\begin{equation}
|\Phi_1>=|\phi_d(23)>|{\bf q_0}>_1 \: . \label{eq:A.2}
\end{equation}
The scattering state antisymmetrized in the two protons is then
explicitly given by
\begin{equation}
\Psi^{(+)} = ( 1-P_{12})\Psi_1^{(+)} \label{eq:A.3}
\end{equation}
and enters the matrix element for elastic p-d scattering as
\begin{equation}
M=<\Phi_{\bf q'}|(V_{12}+V_{13})|\Psi^{(+)}>=
<\Phi_{\bf q'}|(V_{12}+V_{13})(1-P_{12})|\Psi_1^{(+)}>. \label{eq:A.4}
\end{equation}

\noindent
Here $|\Phi_{\bf q'}\rangle$ is defined analogously to $|\Phi_1\rangle$.
Attempts to derive Lippmann-Schwinger equations for the
fully antisymmetrized scattering state given in 
Eq.~(\ref{eq:A.3}) did not lead to 
expressions, which are applicable in practice \cite{PT}.

As an aside, we would like to mention a fact, which may not 
generally be known.
The fully antisymmetrized state $\Psi^{(+)}$ obeys the same Lippmann
Schwinger equation as $\Psi_1^{(+)}$, namely
\begin{equation}
\Psi_1^{(+)} = \Phi_1 + G_1 ( V_{13}+V_{12})\Psi_1^{(+)}. \label{eq:A.5}
\end{equation}
The simple reason is that due to Lippmann identities \cite{Lip},
$\Psi_2^{(+)}=P_{12}\Psi_1^{+}$ obeys the homogeneous version of 
Eq.~(\ref{eq:A.5}). The same is true for a target
composed of a general number of A nucleons (N neutron and Z protons). 
This simply reflects  the fact
that Eq.~(\ref{eq:A.5}) does not uniquely define the scattering state.
Using Eq.~(\ref{eq:A.5}) 
for the state $\Psi^{(+)}$ is only meaningful, if in the course 
of the solution antisymmetrization is always imposed, which is not easily
implemented.

Here we do not want to follow these thoughts. 
Instead we apply the operator $P_{12}$ in Eq.~(\ref{eq:A.4})
to the left. This results in
\begin{eqnarray}
M &=& <\Phi_{\bf q'}|(V_{12}+V_{13})|\Psi_1^{(+)}> - <\phi_{\bf q'}|
(V_{12}+V_{23})|\Psi_1^{(+)}> \nonumber \\
& = & <\Phi_{\bf q'}|(1-P_{12})V_{12}|
\Psi_1^{(+)}>+<\Phi_{\bf q'}|V_{13}|\Psi_1^{(+)}>-<\Phi_{\bf q'}|
P_{12}V_{23}|\Psi_1^{(+)}>. \label{eq:A.6}
\end{eqnarray}
The generalization to a number of A target particles (N neutrons
and Z protons) can be easily obtained in a similar fashion.

\hspace*{10mm}
The first term in the second equality of Eq.~(\ref{eq:A.6}) 
imposes antisymmetry among the projectile 
 target protons by antisymmetrizing the final channel state.
The second term includes the proton-neutron interaction, which requires
no antisymmetrization. The last term is an exchange term different from
 the first one, since the interaction $V_{23}$ acts inside the initial 
target state. As a note, this last term is always
considered of higher order in 
applications of the spectator expansion model \cite{med2}, and is
thus neglected when only the first order term is considered. 

\hspace*{10mm}
Since the physical interaction $V_{12}$ between the two 
protons has to be represented
under all circumstances in physically allowed states, 
i.e. states which are antisymmetric 
under exchanges of (12), 
the operation $(1-P_{12})$ provides just a factor of 2.
Neglecting the last term in Eq.~(\ref{eq:A.6}) we are left with
\begin{equation}
M\rightarrow 2<\Phi_{\bf q'}|V_{12}|\Psi_1^{(+)}> + <\Phi_{\bf q'}|
V_{13}|\Psi_1^{(+)}>.  \label{eq:A.7}
\end{equation}

\hspace*{10mm}
In the following we  sketch the steps usually
carried out in the spectator expansion.
The Green's operator
$G_1$ in Eq.~(\ref{eq:A.5}) is projected onto a part propagating in the 
deuteron target state only, $G_1^b$, and a remainder
\begin{equation}
\Psi_1^{(+)} = \Phi_1 + G_1^b ( V_{13}+V_{12}) \Psi_1^{(+)}+
 G_1^c (V_{13}+V_{12})\Psi_1^{(+)}. \label{eq:A.8}
\end{equation}
By introducing
\begin{equation}
\chi_{\bf q}^{(+)} = \Phi_{\bf q}+G_1^c ( V_{13}+V_{12}) \chi_{\bf q}^{(+)},
\label{eq:A.9}
\end{equation}
one finds together with 
$G_1^b = \int d^3 q'|\Phi_{\bf q'}>g_0({\bf q'})<\Phi_{\bf q}|$
\begin{equation}
\Psi_1^{(+)} = \chi_{\bf q_0}^{(+)} + \int d^3 q' \chi_{\bf q'}^{(+)}
g_0({\bf q'}) <\Phi_{\bf q'}|(V_{13}+V_{12})|\Psi_1^{(+)}>. \label{eq:A.10}
\end{equation}

\noindent
In view of Eq.~(\ref{eq:A.7}) it appears to be inevitable to 
treat the actions of $V_{12}$ and $V_{13}$ separately 
and  work with two coupled equations referring to 
the proton
and neutron parts of M, respectively.  However, doing so  does not
result in a single optical potential, which can serve as driving term for
a one-body Lippmann-Schwinger equation for the elastic amplitude.

\hspace*{10mm}
From Eq.~(\ref{eq:A.10}) one can derive two coupled equations for
the matrix elements involving $V_{12}$ and $V_{13}$:
\begin{eqnarray}
<\Phi_{\bf q}|V_{12}|\Psi_1^{(+)}>&=&<\Phi_{\bf q}|V_{12}|
\chi_{\bf q_0}^{(+)}> \nonumber \\ 
&+&\int d^3 q'<\Phi_{\bf q}|V_{12}|\chi_{\bf q'}^{(+)}>g_0({\bf q'})
<\Phi_{\bf q'}|(V_{12}+V_{13})|\Psi_1^{(+)}>  \nonumber \\
<\Phi_{\bf q}|V_{13}|\Psi_1^{(+)}>&=&<\Phi_{\bf q}|V_{13}|
\chi_{\bf q_0}^{(+)}> \nonumber \\
&+&\int d^3 q'<\Phi_{\bf q}|V_{13}|\chi_{\bf q'}^{(+)}> g_0({\bf q'})
<\Phi_{\bf q'}|(V_{12}+V_{13})|\Psi_1^{(+)}> \label{eq:A.11}
\end{eqnarray}

\noindent
The next step is to evaluate Eq.~(\ref{eq:A.9}). If one defines
\begin{equation}
V_{1i}\chi_{\bf q}^{(+)} \equiv T_i |\Phi_{\bf q}> ,
\label{eq:A.12}
\end{equation} 
where $i=1,2$, one obtains for Eq.~(\ref{eq:A.9})
\begin{equation}
T_i \Phi_{\bf q} = V_{1i}\Phi_{\bf q} + V_{1i} G_1^c \sum_{j=2,3} 
T_j \Phi_{\bf q}. \label{eq:A.13}
\end{equation} 

\noindent
Collecting all $T_i \Phi_{\bf q} $'s on the left side and inverting
the result yields
\begin{equation}
T_i \Phi_{\bf q}=\tau_i \Phi_{\bf q}+\tau_i G_1^c T_j\Phi_{\bf q}
 \;\;\;\; (j\ne i) , \label{eq:A.14}
\end{equation}
where the operators $\tau_i$ obey 
\begin{equation}
\tau_i = V_{1i}+V_{1i} G_1^c \tau_i \: . \label{eq:A.15}
\end{equation}
\noindent
This result corresponds to the familiar Watson expansion \cite{Watson}.
In first order one neglects the second 
part in Eq.~(\ref{eq:A.14}),
which  describes the consecutive scattering of the projectile 
proton on both target nucleons.
In this first order approximation, the only one which has 
been applied in practice, one has
\begin{equation}
T_i \Phi_{\bf q} \rightarrow \tau_i\Phi_{\bf q}, \label{eq:A.16}
\end{equation}
and the optical potentials for pp and np scattering are
given by
\begin{eqnarray}
<\Phi_{\bf q}|V_{12}|\chi_{\bf q'}^{(+)}> &\rightarrow & <\Phi_{\bf q}|
\tau_2|\Phi_ {\bf q'}>  \\ \nonumber
<\Phi_{\bf q}|V_{13}|\chi_{\bf q'}^{(+)}>&\rightarrow & <\Phi_{\bf q}|
\tau_3|\Phi_ {\bf q'}>. \label{eq:A.18}
\end{eqnarray}

\noindent
Denoting the two transition amplitudes in Eq.~(\ref{eq:A.11}) for
simplicity by $\bar Z_2 \equiv \langle \Phi_{\bf q}|V_{12}| 
\chi_{\bf q_0}^{(+)} \rangle$ and 
$\bar Z_3 \equiv \langle \Phi_{\bf q}|V_{13}| \chi_{\bf q_0}^{(+)}\rangle$
and using a matrix notation, indicated by the
`bar'-operators,  Eq.~(\ref{eq:A.11}) takes the form
\begin{eqnarray}
\bar Z_2 &=& \bar \tau_2 + \bar \tau_2 \bar g_0 ( \bar Z_2 + \bar Z_3)
   \nonumber \\
\bar Z_3 &=& \bar \tau_3 + \bar \tau_3 \bar g_0 ( \bar Z_2 + \bar Z_3). 
\label{eq:A.20}
\end{eqnarray}

\noindent
It remains to solve Eq.~(\ref{eq:A.15}), which takes the form 
\begin{equation}
\tau_i = V_{1i} + V_{1i} G_1 \tau_i - V_{1i} G_1^b \tau_i \: .
\label{eq:A.21}
\end{equation}
\noindent
The propagator $G_1$ still contains the interaction of the
struck target nucleon $i$ with the other target nucleon.
The simplest way to solve  Eq.~(\ref{eq:A.21}) would be to replace 
\begin{equation}
G_1\rightarrow G_0 \: . \label{eq:A.22}
\end{equation}
As an aside, expanding $G_1 = G_0 +G_0 V_{23}G_1$ would lead to 
`medium corrections' given by the potential $V_{23}$. However,
for the clarity of presentation, 
we do not want to pursue this further. 
In the approximation introduced by Eq.~(\ref{eq:A.22}) and 
introducing the free NN t-matrix $t_{1i}$ 
Eq.~(\ref{eq:A.21})  simplifies to
\begin{equation}
\tau_i = t_{1i} - t_{1i} G_1^b \tau_i  \: .\label{eq:A.23}
\end{equation}
\noindent
Sandwiched between channel states this reads in matrix notation
\begin{equation}
\bar \tau_i = \bar t_{1i} - \bar t_{1i} \bar g_0 \bar \tau_{1i}=
(1 + \bar t_{1i} \bar g_0)^{-1}\bar t_{1i} \: .
\label{eq:A.24}
\end{equation}

\hspace*{10mm}
If one keeps the interaction $V_{23}$ in $G_1$ one faces a 
three-body problem, which has never been solved
correctly in that context. The work in \cite{med1}
 provides an approximate 
treatment of the interaction of the struck target nucleon with 
the remainder nucleons treated through a mean field interaction. 
With the three-body technology  available today, 
a correct treatment appears to be feasible and  worthwhile.

\hspace*{10mm}
Continuing with the approximation of Eq.~(\ref{eq:A.22}) and inserting 
the expression of Eq.~(\ref{eq:A.24}) into Eqs.~(\ref{eq:A.20})
one finds
\begin{eqnarray}
\bar Z_2 &=& \bar t_{12} + \bar t_{12} \bar g_0 \bar Z_3  \\ \nonumber
\bar Z_3 &=& \bar t_{13} + \bar t_{13} \bar g_0 \bar Z_2 . \label{eq:A.26}
\end{eqnarray}

\noindent
After a partial wave decomposition, 
these are simple one-dimensional integral equations.
Their on-shell solution provides according to Eq.~(\ref{eq:A.7})
the physical transition amplitude
\begin{equation}
M=2 \bar Z_2 + \bar Z_3 . \label{eq:A.27}
\end{equation}
\noindent
The optical potentials occurring in Eqs.~(\ref{eq:A.20})
\begin{equation}
\bar \tau_{2,3}\equiv <\Phi_{\bf q}|\tau_{2,3}|\Phi_{\bf q'}>\label{eq:A.28}
\end{equation}
are the full-folding expressions of the NN t-matrices modified 
according to Eq.~(\ref{eq:A.23}) 
and integrated over the single nucleon density matrix generated 
from the deuteron wave function. 
The NN t-matrix contains the free 3N propagator 
$G_0=(E+i\varepsilon - \frac {p^2}{m}-\frac{3}{4m} q^2)$
and thus in addition to the
kinetic energy of relative motion within the pair, $\frac {p^2}{m}$, 
also the kinetic energy of the pair as a whole in relation to the 
third particle, $\frac{3}{4m} q^2$. 
In Ref.~\cite{brieva} the kinetic energy of the pair is treated within
the full-folding model, while in Refs.~\cite{ffrc,ffce} this dependence
is frozen into a constant.
At this point we do not want to discuss the rationale for the two
different treatments, but rather refer to the literature.

\hspace*{10mm}
It is appealing to compare the expressions of Eq.~(\ref{eq:2.28}) 
for the optical potentials in the 
Faddeev treatment to the one of Eq.~(\ref{eq:A.24}) in
the spectator expansion. However, this is  quite 
hampered.
In Eq.~(\ref{eq:2.28}) a NN t-matrix $t_c$ occurs which 
is modified by a deuteron state for the
same pair of nucleons as for the t-matrix, whereas in Eq.~(\ref{eq:A.24}) 
the deuteron is composed
by a different pair of nucleons. Furthermore, in Eq.~(\ref{eq:2.28}) there 
is the correction term
$(1-G_0g_0^{-1})$ and the factor 2 resulting from $(1-P_{1j})$. In the 
spectator expansion scheme the factor of 2 enters only in the final 
form of Eq.~(\ref{eq:A.27}) and not in the
integral equations (\ref{eq:A.18}),
which are set up for distinguishable particles.

\hspace*{10mm}
In view of all these differences a numerical study would be of 
great interest to clarify at least
in the three-body context the validity of the spectator expansion 
against the systematic Faddeev approach.

\section{The transition potentials}

\hspace*{10mm}
The transition potentials entering the coupled integral equations for
the p-$^3$He and d-d channels are defined in Eq.~(\ref{eq:3.62}). After
considering the amplitudes $\psi_i^c$ and $\psi_i^d$ up to second order
in $T^c$ and $\tilde T^c$ we find for the optical potential for 
elastic p-$^3$He scattering, $V_{\bf u'u}$, the expression given in
Eq.~(\ref{eq:3.68}). In a similar fashion we obtain for the remaining
transition potentials the following: \\
The d-d to d-d transition potential is given by
\begin{eqnarray}
V_{\bf v'v}&=& <\Phi^{dd}_{\bf v'}| V_{12} \tilde P (1-P_{34})
   G_0 T^c P | \Phi^F_{dd,\bf v}> \nonumber \\
&+& <\Phi^{dd}_{\bf v'}| V_{12} \tilde P (1-P_{34}) G_0 T^c P(-P_{34})
      G_0 T^c P | \Phi^F_{dd,\bf v}> \label{eq:B.1}
\end{eqnarray}
The d-d to p-$^3$He transition potential is given by
\begin{eqnarray}
V_{\bf u'v}&=& <\Phi_{\bf u'}| V_{12} P|\Phi^F_{dd,\bf v}> 
  +<\Phi_{\bf u'}| V_{12} P(-P_{34}) G_0 T^c P|\Phi^F_{dd,\bf v}> 
    \nonumber \\
&+& <\Phi_{\bf u'}| V_{12} P(-P_{34}) G_0 T^c P(-P_{34})
  G_0 T^c P |\Phi^F_{dd,\bf v}>  \nonumber \\
&+& <\Phi_{\bf u'}| V_{12} PG_0 \tilde T^c \tilde P (1-P_{34})
    G_0 T^c P |\Phi^F_{dd,\bf v}>  \label{eq:B.2}
\end{eqnarray}
The  p-$^3$He to d-d transition potential is given by
\begin{eqnarray}
V_{\bf v'u}&=&<\Phi^{dd}_{\bf v'}|V_{12} \tilde P(1-P_{34})|\Phi^F_{\bf u}>
 + <\Phi^{dd}_{\bf v'}|V_{12} \tilde P (1-P_{34}) G_0 T^c P(-P_{34})|
   \Phi^F_{\bf u}> \nonumber \\
&+& <\Phi^{dd}_{\bf v'}|V_{12} \tilde P(1-P_{34}) G_0 T^c P(-P_{34})
   G_0 T^c P(-P_{34}) | \Phi^F_{\bf u}> \nonumber \\
&+&<\Phi^{dd}_{\bf v'}| V_{12} \tilde P(1-P_{34}) G_0 T^c P G_0 
\tilde T^c \tilde P (1-P_{34}) |\Phi^F_{\bf u}>    \label{eq:B.3}
\end{eqnarray}

\noindent
After introducing for $G_0 T^c$ and $G_0 \tilde T^c$
 the approximations given in
Eqs.~(\ref{eq:3.74}) and (\ref{eq:3.76}) as well as considering only
terms up to second order in $t_{12}$, we obtain for the transition
potentials:
\begin{eqnarray}
V_{\bf v'v} &=& <\Phi^{dd}_{\bf v'} | V_{12} \tilde P(1-P_{34}) \Lambda
   G_0 t_{12} P 
+V_{12} \tilde P(1-P_{34})\Lambda G_0 t_{12}P G_0 t_{12}P|\Phi^F_{dd,\bf v}>
 \nonumber \\
&+&<\Phi^{dd}_{\bf v'}|V_{12} \tilde P(1-P_{34}) \Lambda G_0 t_{12}
  P(-P_{34}) \Lambda G_0 t_{12} P |\Phi^F_{dd,\bf v}> . \label{eq:B.4}
\end{eqnarray}
\begin{eqnarray}
V_{\bf u'v} &=& <\Phi_{\bf u'} | V_{12} P | \Phi^F_{dd,\bf v}> \nonumber \\
&-&<\Phi_{\bf u'}|V_{12} P P_{34} \Lambda G_0 t_{12} P
   + V_{12} P P_{34} \Lambda G_0 t_{12} P G_0 t_{12}P|\Phi^F_{dd,\bf v}>
  \nonumber \\
&+&<\Phi_{\bf u'}|V_{12} P P_{34} \Lambda G_0 t_{12} P P_{34}
  \Lambda G_0 t_{12} P|\Phi^F_{dd,\bf v}> \nonumber \\
&+&<\Phi_{\bf u'}|V_{12} P \Lambda_{dd} G_0 t_{12} 
 \tilde P(1-P_{34}) \Lambda G_0 t_{12} P|\Phi^F_{dd,\bf v}>
\label{eq:B.5}
\end{eqnarray}
and
\begin{eqnarray}
V_{\bf v'u}&=& <\Phi^{dd}_{\bf v'}|V_{12} \tilde P(1-P_{34})|
\Phi^F_{\bf u}>  \nonumber \\
&-& <\Phi^{dd}_{\bf v'}|V_{12} \tilde P (1-P_{34})\Lambda G_0
t_{12} \left[ 1 + P G_0 t_{12} \right] PP_{34}|\Phi^F_{\bf u}> 
  \nonumber \\
&+&<\Phi^{dd}_{\bf v'}|V_{12} \tilde P (1-P_{34})\Lambda G_0 t_{12} PP_{34}
\Lambda G_0 t_{12} PP_{34}|\Phi^F_{\bf u}> \nonumber \\
&+&<\Phi^{dd}_{\bf v'} | V_{12} \tilde P (1-P_{34}) \Lambda G_0
t_{12} P \Lambda_{dd} G_0 t_{12}\tilde P(1-P_{34})|\Phi^F_{\bf u}> .
   \label{eq:B.6}
\end{eqnarray}
After the same simplifications leading to Eq.~(\ref{eq:3.78}) we obtain
the following expressions  for the transition potentials:
\begin{eqnarray}
V_{\bf v'v}&=&<\Phi^{dd}_{\bf v'}|2 V_{12} \tilde P \Lambda G_0 t_{12}
  P |\Phi^F_{dd,\bf v}> \nonumber \\
&+&<\Phi^{dd}_{\bf v'}|2 V_{12} \tilde P \Lambda G_0 t_{12} P
 G_0 t_{12} P | \Phi^F_{dd,\bf v}>  \nonumber \\
&-&<\Phi^{dd}_{\bf v'}|2 V_{12} \tilde P \Lambda G_0 t_{12} PP_{34}
  \Lambda G_0 t_{12} P | \Phi^F_{dd,\bf v}>, \label{eq:B.7}
\end{eqnarray}
\begin{eqnarray}
V_{\bf u'v} &=& <\Phi_{\bf u'} | V_{12} P | \Phi^F_{dd,\bf v}> 
 -<\Phi_{\bf u'}|V_{12} P P_{34} \Lambda G_0 t_{12} P|\Phi^F_{dd,\bf v}>
 \nonumber \\
&-& <\Phi_{\bf u'}|V_{12} P P_{34} \Lambda G_0 t_{12} P [1-P_{34}\Lambda]
  G_0 t_{12} P |\Phi^F_{dd,\bf v}> \nonumber \\
&+& <\Phi_{\bf u'}| 2 V_{12} P \Lambda_{dd} G_0 t_{12} \tilde P
  \Lambda G_0 t_{12} P|\Phi^F_{dd,\bf v}> \label{eq:B.8}
\end{eqnarray}
and
\begin{eqnarray}
V_{\bf v'u}&=& <\Phi^{dd}_{\bf v'}| 2 V_{12} \tilde P |\Phi^F_{\bf u}> 
 - <\Phi^{dd}_{\bf v'}|2 V_{12} \tilde P \Lambda G_0 t_{12}
PP_{34}|\Phi^F_{\bf u}> \nonumber \\
&-& <\Phi^{dd}_{\bf v'} |2  V_{12} \tilde P \Lambda G_0 t_{12} P
[1-P_{34}\Lambda] G_0 t_{12} P P_{34}|\Phi^F_{\bf u}> \nonumber \\
&+& <\Phi^{dd}_{\bf v'} |4 V_{12} \tilde P \Lambda G_0 t_{12} P
 \Lambda_{dd} G_0 t_{12}\tilde P | \Phi^F_{\bf u}>. \label{eq:B.9}
\end{eqnarray}

%----------------------------------------------------

%\pagebreak

\pagebreak
%%%%%%%%%%%%%%%%%%%%%%%%%%%%%%%%%%%%%%%%%%%%%%%%%%%%%%

\noindent
\begin{figure}
\caption{Born series for the four-nucleon breakup process 
  p-$^3$He $\rightarrow$ pppn. Particle 4 is singled out as 
  projectile. For further explanation see text.  \label{fig1}}
\end{figure}

\noindent
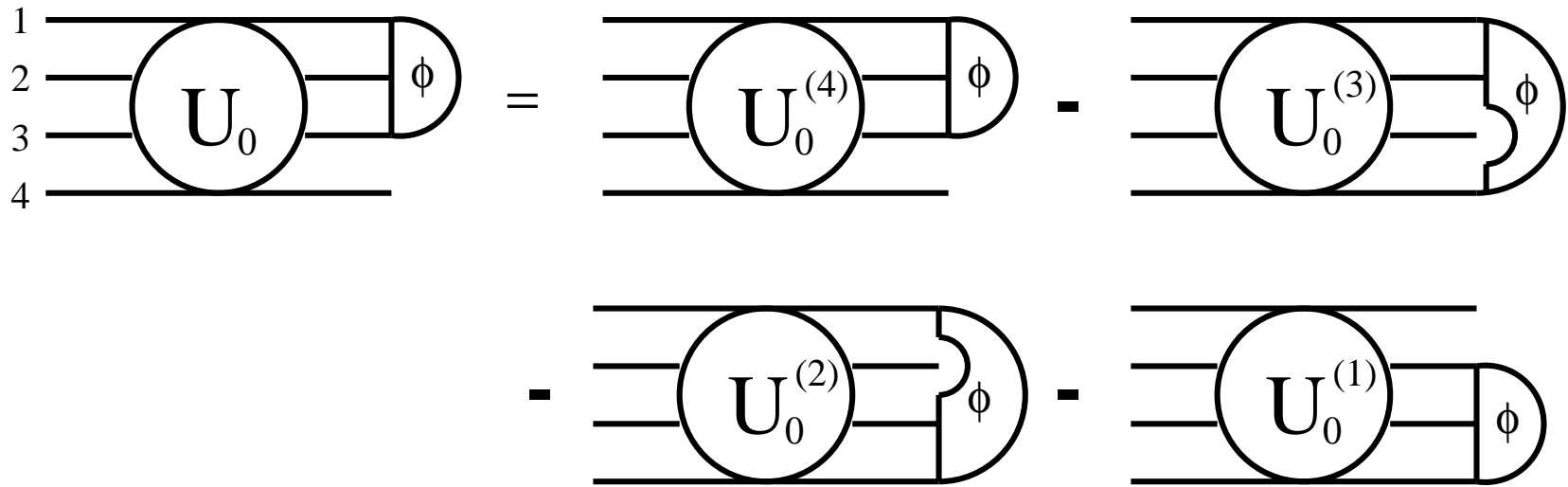
\begin{figure}
\caption{The fully antisymmetrized four-nucleon breakup operator
 $U_0$. Since the target state is assumed to be antisymmetric only
the exchange terms with the projectile have to be considered. This
leads to four terms with each nucleon being the projectile.
\label{fig2}}
\end{figure}

\noindent 
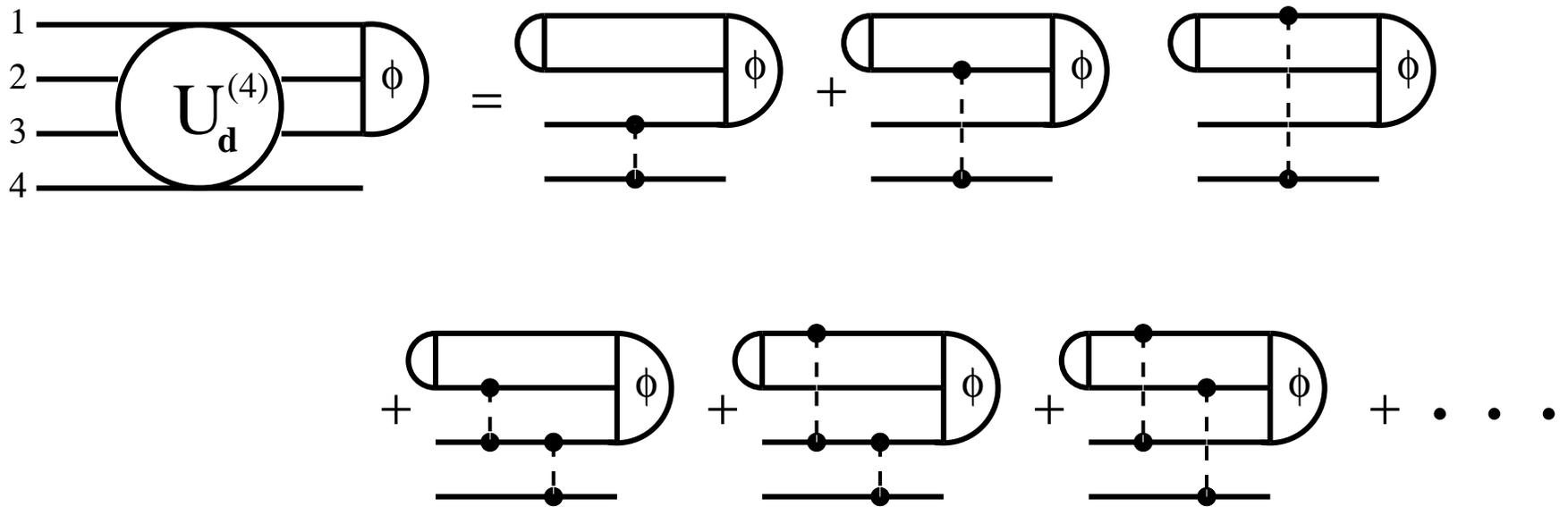
\begin{figure} 
\caption{Born series for the three-body fragmentation process
 p-$^3$He $\rightarrow$ ppd. Particle 4 is singled out as 
  projectile. For further explanation see text.  \label{fig3}}
\end{figure}

\end{document}